\documentclass{aa}  

\usepackage{graphicx}
\usepackage{txfonts}
\usepackage{lipsum}
\usepackage{subcaption}         
\usepackage{lscape}             
\usepackage{placeins}           
\usepackage{tabularx}
\usepackage{array}
\usepackage{booktabs}
\usepackage{tikz}
\usepackage{cleveref}
\usepackage[utf8]{inputenc} 
\DeclareUnicodeCharacter{200B}{}
                                
\begin{document}

   \title{A Machine Learning Framework \protect\\ for Stellar Collision Transient Identification}


   \author{B. X. Hu
          \inst{1}
          \and
          A. Loeb\inst{2}
          }
          
   \institute{Department of Physics, Harvard University, Cambridge, MA, USA\\
             \email{bhu@g.harvard.edu}
            \and Department of Astronomy, Harvard University, Cambridge, MA, USA\\ 
            \email{aloeb@cfa.harvard.edu}
            }

   \date{Received Month Day, Year; accepted Month Day, Year}

 
  \abstract
   {Modern astronomical surveys, such as the Zwicky Transient Facility (ZTF), are capable of detecting thousands of transient events per year, necessitating the use of automated and scalable data analysis techniques. Recent advances in machine learning have enabled the efficient classification and characterization of these transient phenomena.}
   {We aim to develop a fully systematic pipeline to identify candidate stellar collision events in galactic nuclei, which may otherwise be identified as tidal disruption events or other transients. We also seek to validate our simulations by comparing key physical parameters derived from observations and used in modeling these events.} 
   {We generate a comprehensive bank of simulated light curves spanning a range of physical parameters and employ an approximate nearest neighbor algorithm (via the \texttt{annoy} library) to match these with observed ZTF light curves.}
   {Our pipeline is successfully able to associate observed ZTF light curves with simulated events. The resulting estimated parameters, including supermassive black hole masses and ejecta mass, are presented and compared to known values when applicable.}
   {We demonstrate that a systematic, machine learning-based approach can effectively identify and characterize stellar collision candidate events from large-scale transient surveys. This methodology is especially promising for future surveys which will provide us with significantly high volumes of data, such as LSST, where automated, data-intensive analysis will be critical for advancing our understanding of transient astrophysical phenomena.}

   \keywords{supermassive black holes --
                transients --
                machine learning --
                simulation
               }

   \maketitle
   \nolinenumbers

\section{Introduction}\label{introduction}

Machine learning is the branch of artificial intelligence that enables systems to learn patterns from data and improve their performance over time without explicit programming. In recent decades, it has dramatically transformed astrophysics by providing powerful tools to analyze increasingly large and complex datasets. Early applications focused on relatively straightforward classification tasks—such as distinguishing stars, galaxies, and quasars using methods like decision trees and support vector machines—which enabled researchers to efficiently sift through data from surveys such as SDSS and Pan-STARRS~\citep{ball2010, baron2019}. As the volume and complexity of the data grew, the field rapidly adopted more sophisticated techniques, including deep learning and convolutional neural networks, to tackle more advanced challenges such as image recognition and time-series analysis~\citep{dieleman2015,dubuisson2015,dominguezsanchez2018}.

Recent developments in machine learning have further advanced our ability to study time-domain phenomena. Innovative methods—ranging from variational autoencoders to advanced anomaly detection algorithms and probabilistic inference techniques—are now routinely employed to analyze transient events in massive datasets from modern surveys like the Australian Dark Energy Survey \citep[OzDES,][]{yuan2015,childress2017}, and will certainly play a critical role in even larger, upcoming surveys such as the Vera C. Rubin Observatory Legacy Survey of Space and Time (LSST)~\citep{giuseke2017,muthukrishna2019,portillo2020}. These cutting-edge approaches have not only accelerated data processing but have also enabled researchers to uncover previously inaccessible correlations and patterns, providing insight into dynamic events such as tidal disruption events, gravitational lensing, supernovae, gamma-ray bursts, and more. 

Beyond these technical advances, the integration of machine learning with traditional astrophysical modeling has fostered significant interdisciplinary collaboration, bringing together computer scientists and astronomers to tackle challenges that were once considered insurmountable~\citep{fluke2019}. In parallel, there is growing emphasis on improving model interpretability and quantifying uncertainties in predictions—an essential step in ensuring that data-driven insights are physically meaningful and robust when comparing simulation results with observational data~\citep{gal2016,kendall2017}. Furthermore, emerging techniques such as generative adversarial networks (GANs) and variational autoencoders (VAEs) are being deployed to create realistic simulations, estimate parameters, and model and discover astrophysical phenomena, thereby expanding our ability to model complex astrophysical phenomena~\citep{storeyfisher2021,thorne2021,gabbard2022}.

In our work, we leverage the approximate nearest neighbor algorithm to match observed light curves of candidate tidal disruption events (TDEs) from the Zwicky Transient Facility (ZTF) with a comprehensive bank of simulated light curves resulting from collisions of stars in galactic nuclei. This approach reduces the computational complexity inherent in large-scale searches and provides a robust framework for comparing observational data with theoretical models. By integrating advanced machine learning methods with simulation-based studies, our analysis aims to identify subtle patterns and similarities that can bridge the gap between empirical observations and the underlying astrophysical processes.

We studied the topic of stellar collisions in galactic nuclei in our previous works \citet{hu2024a} and \citet{hu2024b}. In \citet{hu2024a}, we calculated the rate of stellar collisions and studied how they varied with parameters such as the supermassive black hole (SMBH) mass $M_{\bullet}$, galactocentric radius $_{\mathrm{gal}}$, and energy of the ejecta $E_{\mathrm{ej}}$. We used the simplified, semi-analytic "radiative zero" approach by Arnett~\citep{arnett1996} to model the explosion light curve and found that while the events could be highly energetic, with luminosities on order of those of supernovae, the light curves resulting from the explosions alone were also expected to decay fairly quickly, which would make them difficult to detect even by modern surveys. 

In \citet{hu2024b}, we then studied the accretion flares that can result from the collisions forming streams of debris that can accrete onto the SMBH. We simulated the post-collision ejecta as $N$ particles in a circular distribution, all of equal mass, starting at the site of the collision and moving around the SMBH along geodesics of spacetime. We adopted the widely-used alpha-disk model~\citep{shakura1973} to calculate how each particle's accretion timescale was affected by the presence of a SMBH's accretion disk. It was found that the post-collision accretion flare can take on a wide variety of appearances based on several factors, including $M_{\bullet}$, $_{\mathrm{gal}}$, and the velocity vector of the ejecta, $\vec{v}_{\mathrm{ej}}$. The light curves resulting from our simulations contained unique observational signatures that we believe can be used to identify and distinguish these events from other astrophysical transients.

Stellar collisions, supernovae, and tidal disruption events are all energetic transients that can have overlapping peak luminosities, yet their underlying physical mechanisms and light curve evolution differ markedly. Supernovae, for example, are powered by the re-ignition of nuclear fusion in a white dwarf or by the core-collapse processes within stars, typically producing relatively uniform light curve profiles and well-established spectral signatures from nucleosynthesis products. In contrast, TDEs occur when a star is disrupted by the tidal forces of a supermassive black hole, leading to a prolonged, gradually decaying flare as the stellar debris is steadily accreted. Stellar collisions in galactic nuclei, as described in our previous works, are distinguished by an initial, brief, yet luminous explosion, followed by an accretion flare whose properties are highly sensitive to various parameters of the system. In addition, both TDEs and the stellar collisions we have described occur only in galactic nuclei, further allowing them to be distinguished from supernovae. By comparing and contrasting these distinct characteristics, we can develop robust criteria for classifying these events. This differentiation is essential for leveraging observational data to accurately identify the astrophysical processes at play and for guiding future modeling and survey strategies.

The structure of this paper is as follows. In section \ref{method}, we will describe our analysis pipeline, including the machine learning method used. In section \ref{results}, we will present the results of fitting to several light curves of interest, identified using this method, as well as parameter estimation. In section \ref{discussion}, we will discuss the implication of our results as well as directions for future work.

\section{Method}\label{method}

The basic structure of our data analysis and parameter estimation pipeline is as follows. We generate a bank of simulated observational data (light curves), each initialized with a different set of physical parameters. We use an astronomical data broker to access observed data from the Zwicky Transient Facility (ZTF). We filter ZTF data to only consider light curves which could possibly originate from TDEs, which we believe our stellar collision events of interest are most likely to be confused with. For each ZTF light curve of interest, we employ the approximate nearest neighbor algorithm to find the simulation which most closely resembles the observed data, and assign a goodness-of-fit score. We select a handful of the best fits to further explore for the purpose of this study. For each of these light curves, we identify the most likely host galaxy, and estimate various physical parameters of interest to identify whether or not these events could be stellar collisions. This full pipeline is described in more detail in this section, and illustrated in Fig. \ref{fig:schematic}. 

We note that while Markov Chain Monte Carlo (MCMC) methods are widely used in astrophysics for parameter estimation~\citep{lewis2002}--with numerous Python implementations available, such as the popular \textit{emcee} package~\citep{foremanmackey2013}--we did not adopt this approach for our study due to the computational expense associated with our simulations. MCMC is a Bayesian technique that efficiently samples the posterior distribution of parameters by constructing a Markov chain whose stationary distribution approximates the true parameter probability distribution~\citep{hastings1970}. In principle, MCMC would allow us to estimate the physical parameters of the stellar collision system directly from an observed light curve. However, because each of our simulations takes several hours to generate a complete light curve, employing MCMC—easily requiring thousands or even tens of thousands of simulation evaluations—would be computationally prohibitive. Consequently, we opted for an approximate nearest neighbor algorithm to rapidly match observed light curves to our pre-computed simulation bank, thus providing a more feasible and efficient pathway for parameter estimation.

We follow the simulation method described in \citet{hu2024b} to comprehensively explore the parameter space describing stellar collisions in galactic nuclei. In our simulations, the supermassive black hole mass, $M_{\bullet}$​, is varied over 8 discrete values from $5\times10^5\,M_{\odot}$ to $10^9\,M_{\odot}$. For each value of $M_{\bullet}$, we also consider 16 discrete initial radial positions, $r_{\mathrm{init}}$, from $50\,r_S$ to $10^9\,r_S$, with $r_S$ the Schwarzschild radius, which in turn determine the magnitude of the relative velocity $\vec{v}_{\mathrm{rel}}=\vec{v}_1-\vec{v}_2$, where $\vec{v}_1$ and $\vec{v}_2$ are the velocities of the two stars colliding. This relative velocity is drawn from a Maxwellian distribution that describes a galaxy's stellar density profile $\rho_{\eta}(r_{\mathrm{gal}})$ provided by \citet{tremaine1994}, 
   \begin{equation}\label{eq:ch3_1}
    \ \rho_{\eta}(r_{\mathrm{gal}})\equiv\frac{\eta}{4\pi}\frac{r_sM_{\mathrm{\mathrm{sph}}}}{r_{\mathrm{gal}}^{3-\eta}(r_s+r_{\mathrm{gal}})^{1+\eta}},
    \end{equation}
    where we adopted the commonly used value $\eta=2$~\cite{hernquist1990}, $M_{\mathrm{sph}}$ is the total mass of the host spheroid, $r_s$ is a distinctive scaling radius, and $r_{\mathrm{gal}}$ is the galactocentric radius, which we also call the initial radius $r_{\mathrm{init}}$ throughout our work. The values of $M_{\mathrm{sph}}$ and $r_s$ are both determined by the value of $M_{\bullet}$ adopted in the specific simulation, as described in \citet{hu2024a}. The orientation of the relative velocity vector is specified by two angles: $\phi$, which spans from $0$ to $2\pi$, and $\theta$, which spans from $0$ to $\pi$. For both $\phi$ and $\theta$, 12 uniformly spaced values are sampled. Altogether, this results in a total of $8\times16\times12\times12=18,432$ simulations, ensuring a thorough exploration of how variations in SMBH mass, initial radial position, and velocity orientation impact the dynamics of stellar collisions and the properties of the subsequent accretion flares. 

    Our simulations are implemented in Python and executed on Harvard’s FAS Research Computing cluster, which allows us to take advantage of high-performance parallel processing. Each simulation is run independently, and the resulting data are stored in HDF5 (h5) files--a format well-suited for managing large, complex datasets due to its efficient storage and rapid I/O capabilities. This setup streamlines our workflow and allows for easy data management and subsequent analysis.

    We acquire observed light curve data from Lasair~\citep{smith2019,williams2024}. Lasair is an alert broker designed to collect and process transient alerts from surveys, built with the ultimate goal of serving transient alerts from the upcoming LSST. In the meantime, its functionality is being prototyped using data from the ZTF. By aggregating, filtering, and distributing alert data in near real-time, Lasair enables researchers to quickly access light curves and other time-domain observations. In our work, we leverage Lasair’s filtering capabilities to efficiently isolate candidate tidal disruption events (TDEs) from the vast ZTF dataset. In addition to Lasair, there exist many alternative brokers such as ANTARES~\citep{narayan2019}, ALeRCE~\citep{forster2021}, and AMPEL~\citep{nordin2019}, all of which also offer alert management and filtering tools. We ultimately chose Lasair for its ease of use and flexible query interface, in particular with respect to TDE candidates, which streamlines the process of acquiring and sifting through observational data for our analysis.

    The Zwicky Transient Facility (ZTF) is a fully-automated, wide-field survey designed to scan and monitor the entire optical Northern sky every two days for transient astrophysical phenomena that rapidly change in brightness. Its camera is consists of 16 CCDs, each of 6,144x6,160 pixels, enabling each exposure to cover an area of 47 square degrees. Building on the legacy of predecessors like the Palomar Transient Factory \citep[PTF,][]{rau2009}, ZTF combines a large field of view with high cadence and rapid image processing to capture a broad range of transient events, such as supernovae, gamma ray bursts, tidal disruption events, and more~\citep{bellm2019}. Since its commissioning in 2018, ZTF has provided a large, high-quality dataset of light curves that is particularly valuable for time-domain studies. Its capability to record high-cadence photometric data makes it well-suited for our work, as it increases the likelihood of detecting brief, rapidly evolving events, such as those arising from stellar collisions in galactic nuclei. Consequently, the extensive and timely data provided by ZTF offers an ideal observational basis for matching observed light curves with our simulation bank, allowing us to test our theoretical models. ZTF is expected to accumulate ten times the amount of data compared to PTF, and serves as a prototype for LSST, which in turn is expected to generate ten times more data than ZTF. 
    
    The Lasair alert broker allows users to create and run SQL queries to return past (or present) events which match a certain set of conditions. We choose to query for TDE candidates because, as discussed in \citet{hu2024a} and \citet{hu2024b}, the fact that both these kinds of events occur in galactic nuclei leads us to believe that stellar collisions could be observed and incorrectly interpreted as TDEs. As long as a sufficiently broad set of criteria is provided in the query, we expect accretion flares from stellar collision events to be captured by criteria commonly associated with searching for TDEs. We adopt a set of criteria that, in addition to hopefully capturing both TDE and stellar collision candidates, also ensure a reasonably high quality (i.e. high signal to noise ratio) for the light curves which we will then analyze and compare to our simulations. The set of criteria we use is describe in Table \ref{tab:lasair_criteria}. 
    
    In addition to requiring the objects to be nuclear, high-quality, and have a reasonably high number of data points, we also require the observation to not have occurred within 10 degrees of galactic latitude, to minimize contamination from the high density of stars, increased dust extinction, and other variable sources near the galactic plane. We also require the mean color or latest color to be blue, because TDEs are typically characterized by high effective temperatures that produce blue, UV-dominated emission, so selecting for sources with consistently blue colors helps isolate events with spectral properties expected from TDEs~\citep{gezari2021}. Because the accretion mechanism for stellar collision ejecta is likely very similar, we keep this criteria for our events. TDEs are more frequently associated with quiescent or post-starburst galaxies that have older stellar populations and exhibit redder optical colors. This selection helps reduce contamination from star-forming, bluer hosts and increases the likelihood that the transient is extragalactic and consistent with known TDE host properties~\citep{french2016}. Observable emission from stellar collisions in galactic nuclei is expected to occur preferentially in older, redder hosts. This is because the dense stellar environments conducive to such collisions are often found in the central regions of early-type or quiescent galaxies, where the stellar populations are typically evolved and redder in color~\citep{cote2006}. Consequently, similar to TDE candidates, the selection of red host galaxies can serve as an effective criterion for isolating events that arise in these mature, high-density environments. 
    
    Once the ZTF light curve data are acquired as apparent magnitude versus time in the observer frame, we transform our simulation data, which are in the form of luminosity versus time in rest frame, such that it can be directly compared to the observed data. For each light curve, we use the magnitude of its reference source as well as its redshift (either directly known or assumed to be a modest value) to transform each simulated light curve into apparent magnitude versus time. In addition, there are two free parameters introduced to allow for scaling on both the x- and y-axes. A free parameter is used to shift the time axis, allowing for an offset between the onset of observations and the initialization of the simulation, so we do not assume that the first observation by ZTF aligns exactly with the start of the simulation. An addition parameter is used to scale the luminosity linearly with the mass of the ejecta. After parameter estimation, these assumptions—both the linear scaling of luminosity with ejecta mass and the temporal offset—can be checked more thoroughly to ensure that they do not lead to any unphysical parameters or assumptions. 

Next, we turn to the approximate nearest neighbor algorithm to match each observed light curve to a simulation from our bank. The nearest neighbor (NN) search is a classic optimization problem that appears widely throughout not only computer science, but also in various scientific domains and beyond. The problem can be stated as: given a set of points $S\subseteq\mathbb{R}^d$ and a query point $q\in\mathbb{R}^d$, the goal is to identify the point $p\in S$ that minimizes the distance function $d(p,q)$. The distance function, also known as the dissimilarity function, is used to express closeness between the points: the larger the distance or dissimilarity between the points, the larger the function will evaluate to. This problem is relatively straightforward in low-dimensional spaces, where data structures such as $k$-dimensional ($k$-d) trees\footnote{$k$-d trees are space-partitioning binary trees that are constructed by recursively splitting the dataset with hyperplanes that are perpendicular to one of the coordinate axes. At each level, the splitting axis typically cycles through the available dimensions, dividing the data into two subsets. This structure facilitates rapid range queries and nearest neighbor searches by eliminating large portions of the search space.~\cite{bentley1975}} can efficiently handle queries by recursively partitioning the space. However, as the dimension $d$ grows larger, exact nearest neighbor searches become prohibitively expensive in both time and memory (an effect known as the curse of dimensionality).

To mitigate this issue, researchers introduced the concept of approximate nearest neighbor (ANN) searches, which, as the name suggests, relax the strict requirement of identifying the absolute closest neighbor. Instead, ANN algorithms seek to find a point that is “close enough” to the query with high probability, drastically reducing computational overhead. ANN algorithms are discussed more in-depth in Appendix \ref{app:A}.

After using \texttt{annoy} to identify the approximate nearest neighbor simulation and assigning a corresponding goodness-of-fit score, we further examine several of the most promising fits between simulations and candidate ZTF light curves. For each transient, we attempt to use known information to extract the physical properties of the associated host galaxy to assess the feasibility of a stellar collision event. We then analyze the best-fit simulation to determine if any of its input parameters—such as the mass of the SMBH—offer additional insights or suggest further hypotheses regarding the observed phenomena.

In summary, we have developed a pipeline that integrates a simulation bank of stellar collision light curves with an approximate nearest neighbor search--implemented via \texttt{annoy}--to rapidly and systematically identify candidate matches from ZTF observations which have characteristics matching those of the stellar collision events we are interested in. This methodology allows for the efficient selection of promising transient events, with each match assigned a corresponding goodness-of-fit score and supplemented by the host galaxy's known properties to assess the likelihood of a stellar collision scenario. By filtering ZTF data through the Lasair alert broker and employing stringent selection criteria, our approach ensures that only high-quality transient events are advanced for further study. The detailed parameter estimation and subsequent physical interpretation of these candidates are presented in the following sections.

\usetikzlibrary{arrows.meta,
                calc, chains,
                quotes,
                positioning,
                shapes.geometric}
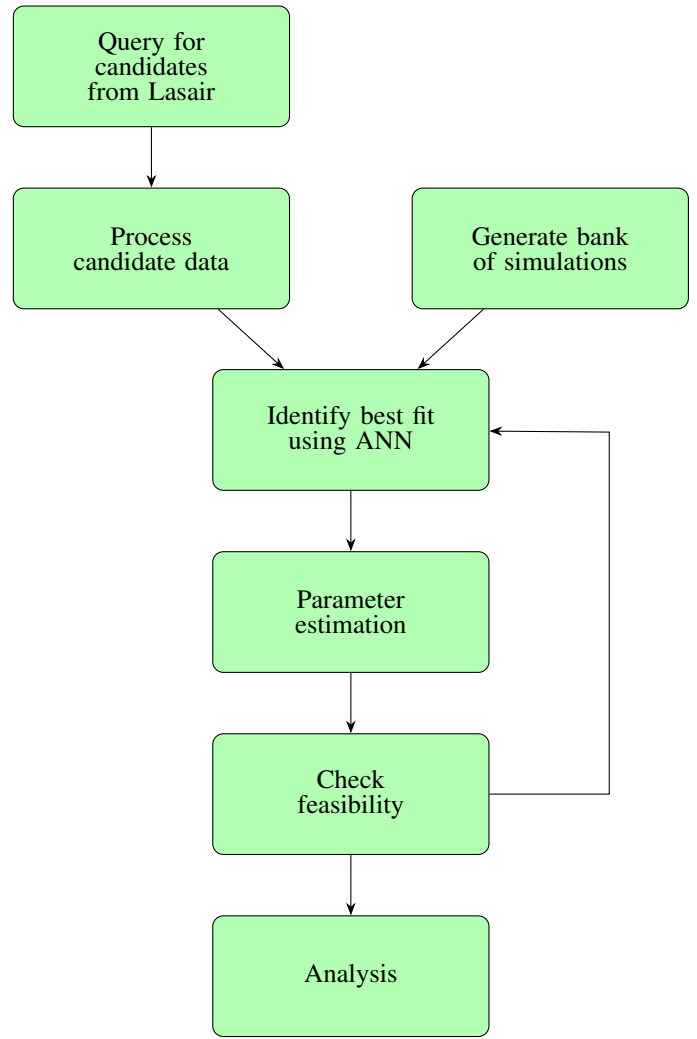
\begin{figure}[p]

\vspace*{\fill}
\centering
\begin{tikzpicture}[
	node distance = 8mm and 16mm,
	start chain = A going below,
	base/.style = {draw, rounded corners, text width=34mm, minimum height=16mm,
		align=center, on chain=A},
	startstop/.style = {base, rectangle, fill=green!30},
	process/.style = {base, rectangle, fill=green!30},
	io/.style = {base, rectangle, fill=green!30},
	decision/.style = {base, rectangle, fill=green!30},
	every edge quotes/.style = {auto=right}]
	]
	
	\node [process] (A1) {Query for \\[-0.35ex]
		candidates \\[-0.35ex]
		from Lasair}; 
	\node [process] (A2) {Process \\[-0.35ex]
		candidate data};
	
	\node [process, right=of A2] (B1) {Generate bank \\[-0.35ex]
		of simulations};
		
	\coordinate (Mid) at ($(A2.east)!0.5!(B1.west)$);
	
	\node [process, below=16mm of Mid] (C1) {Identify best fit \\[-0.35ex]
		using ANN}; 
	\node [process, below=of C1] (C2) {Parameter \\[-0.35ex]
		estimation};
	\node [process, below=of C2] (C3) {Check \\[-0.35ex]
		feasibility};
	\node [process, below=of C3] (C4) {Analysis};

	\draw [arrows=-Stealth] 
	(A1) edge[""] (A2)
	(A2) edge[""] (C1)
	(B1) edge[""] (C1)
	(C1) edge[""] (C2)
	(C2) edge[""] (C3)
	(C3) edge[""] (C4);
	
	\coordinate (B) at ($(C3)+(34mm,0)$); 
	\coordinate (C) at ($(B)+(0,48mm)$); 
	\draw[arrows=-Stealth] (C3) -- (B) -- (C) -- (C1);

\end{tikzpicture}
\vspace*{\fill}
\caption{A schematic illustration of our data analysis and parameter estimation pipeline. We first query for candidate tidal disruption events (TDEs) and stellar collisions from the Lasair broker and process the resulting light curve data. A bank of simulated light curves is generated over a broad parameter space. Each observed light curve candidate is matched to its nearest simulation using an approximate nearest neighbor algorithm. We then perform parameter estimation and verify whether the results are physically plausible. Finally, we assess the feasibility of each candidate as a potential stellar collision event before proceeding to a detailed analysis.}
\label{fig:schematic}

\end{figure}

\section{Results}\label{results}

\begin{figure}[htbp]
   \centering
   \makebox[\textwidth][c]{%
    \includegraphics[width=1\textwidth]{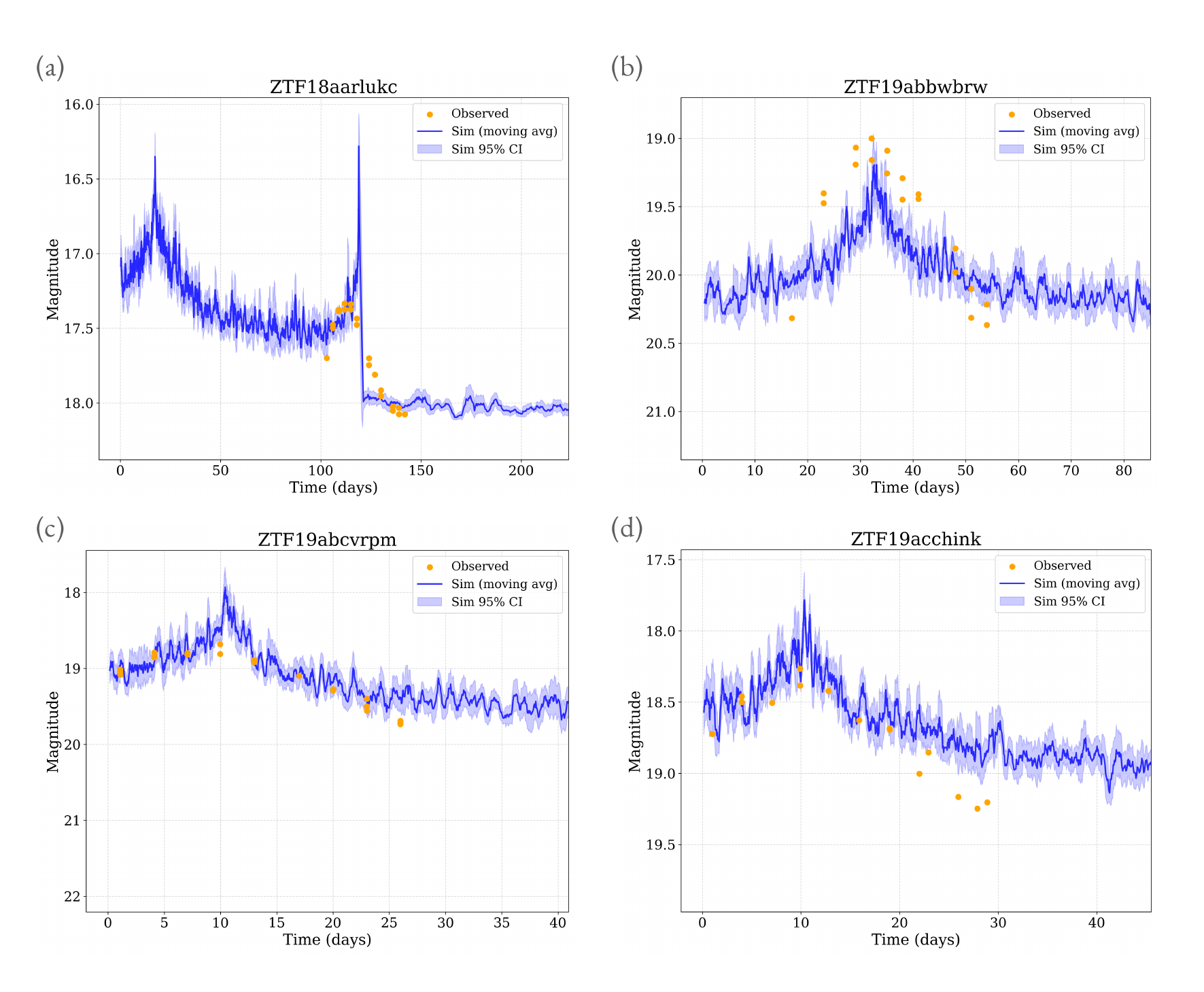}
  }
      \caption{Light curves for four transients: (a) ZTF18aarlukc, (b) ZTF19abbwbrw, (c) ZTF19abcvrpm, (d) ZTF19acchink. In each panel, the observed data from ZTF (orange points) are plotted as apparent magnitude versus time (in days, observer frame); a constant time offset has been applied to improve alignment with the simulation. The blue line shows the 10‑day moving average of the simulation, while the light blue shaded area denotes the corresponding 95\% confidence interval of the simulated data, calculated from the standard error of the simulation magnitudes over the 10-day windows and assuming a normal distribution. The physical parameters used in the simulations—as well as the corresponding known system properties—are summarized in Table \ref{tab:params}.}
         \label{fig:ch3_lcs_1}
   \end{figure}

\begin{figure}[htbp]
   \centering
   \makebox[\textwidth][c]{%
    \includegraphics[width=1\textwidth]{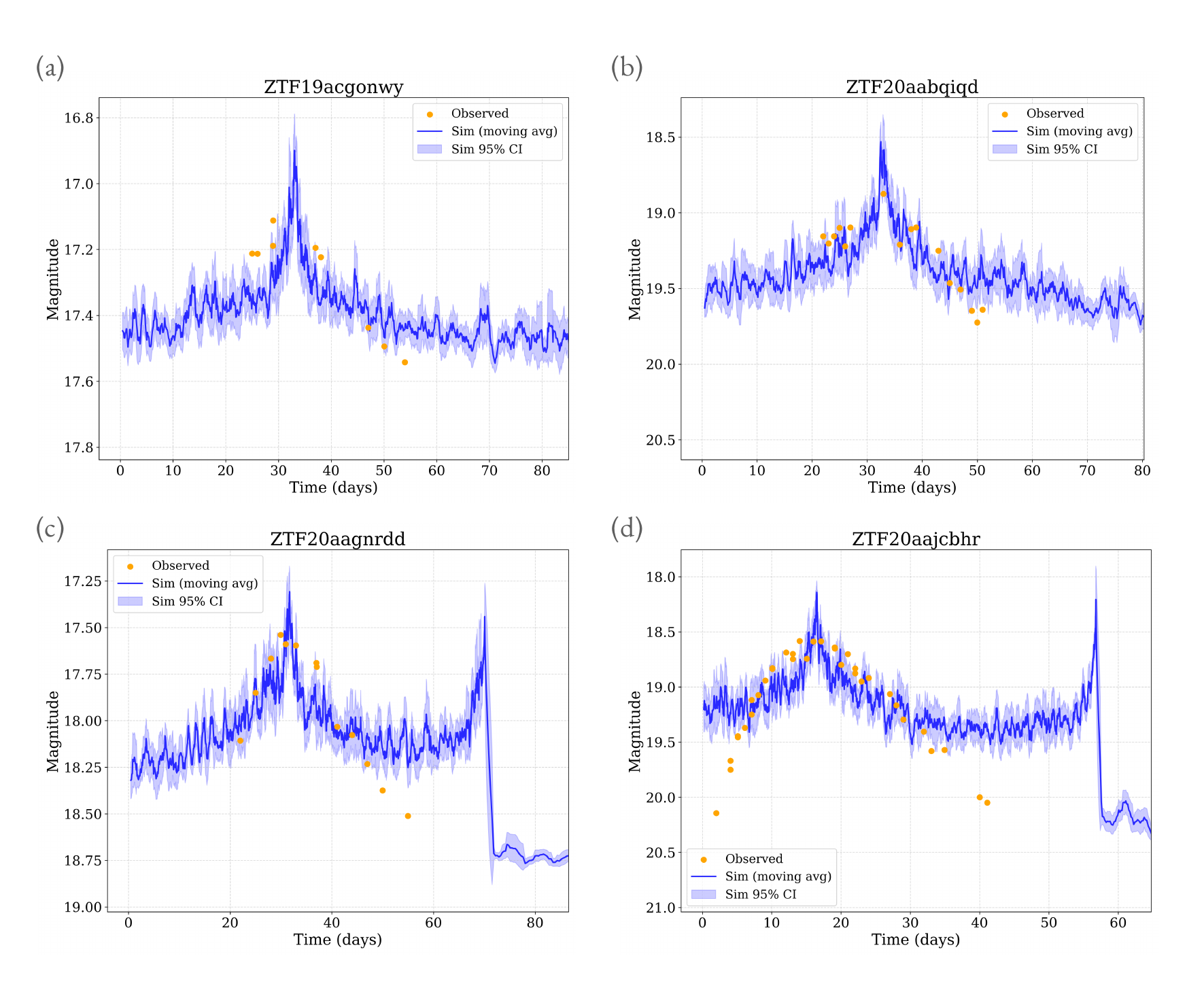}
  }
      \caption{Light curves for four transients: (a) ZTF19acgonwy, (b) ZTF20aabqiqd, (c) ZTF20aagnrdd, (d) ZTF20aajcbhr. In each panel, the observed data from ZTF (orange points) are plotted as apparent magnitude versus time (in days, observer frame); a constant time offset has been applied to improve alignment with the simulation. The blue line shows the 10‑day moving average of the simulation, while the light blue shaded area denotes the corresponding 95\% confidence interval of the simulated data, calculated from the standard error of the simulation magnitudes over the 10-day windows and assuming a normal distribution. The physical parameters used in the simulations—as well as the corresponding known system properties—are summarized in Table \ref{tab:params}.}         \label{fig:ch3_lcs_2}
   \end{figure}

\begin{figure}[htbp]
   \centering
   \makebox[\textwidth][c]{%
    \includegraphics[width=1\textwidth]{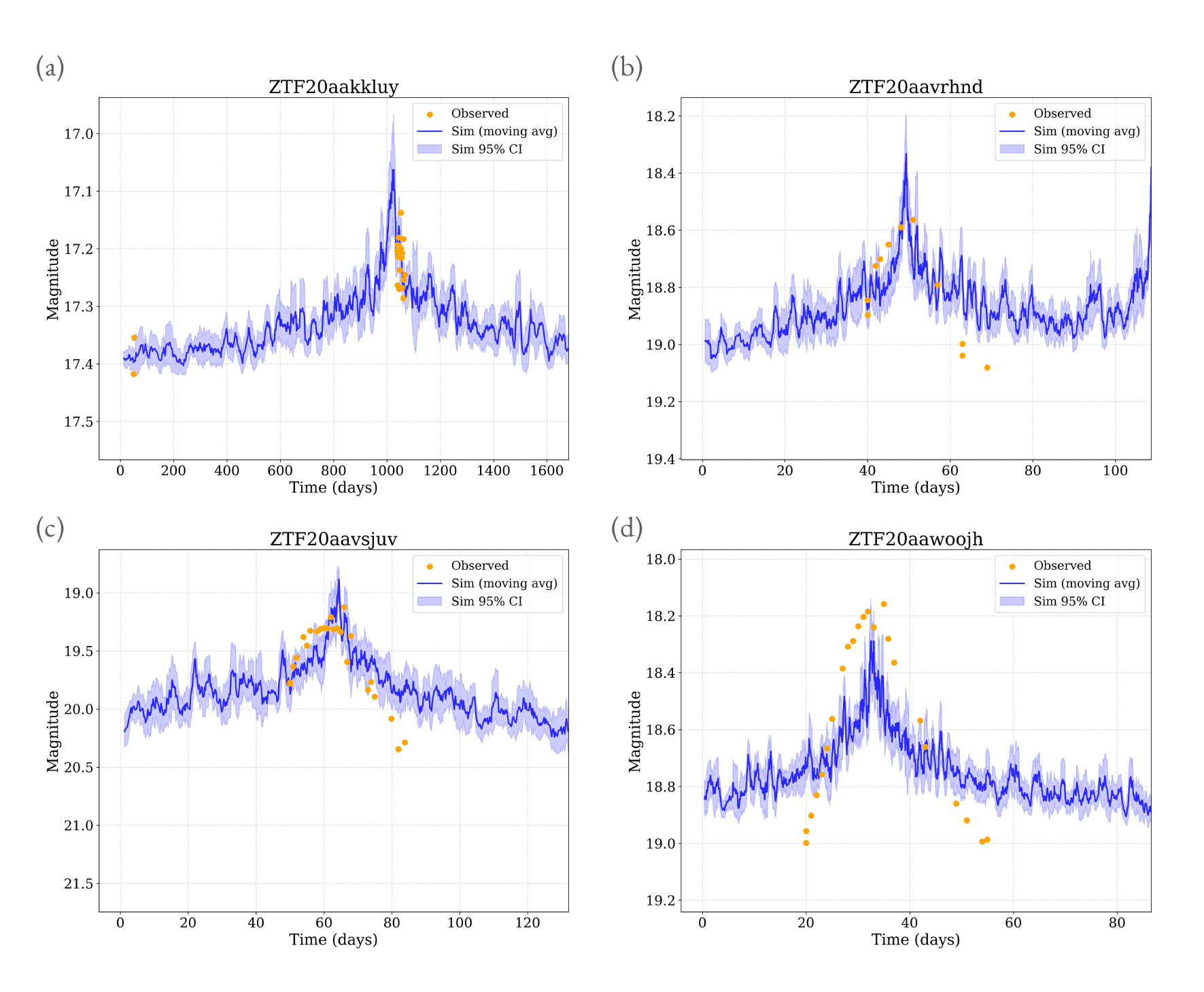}
  }
      \caption{Light curves for four transients: (a) ZTF20aakkluy, (b) ZTF20aavrhnd, (c) ZTF20aavsjuv, (d) ZTF20aawoojh. In each panel, the observed data from ZTF (orange points) are plotted as apparent magnitude versus time (in days, observer frame); a constant time offset has been applied to improve alignment with the simulation. The blue line shows the 10‑day moving average of the simulation, while the light blue shaded area denotes the corresponding 95\% confidence interval of the simulated data, calculated from the standard error of the simulation magnitudes over the 10-day windows and assuming a normal distribution. The physical parameters used in the simulations—as well as the corresponding known system properties—are summarized in Table \ref{tab:params}.}
         \label{fig:ch3_lcs_3}
   \end{figure}

\begin{figure}[htbp]
   \centering
   \makebox[\textwidth][c]{%
    \includegraphics[width=1\textwidth]{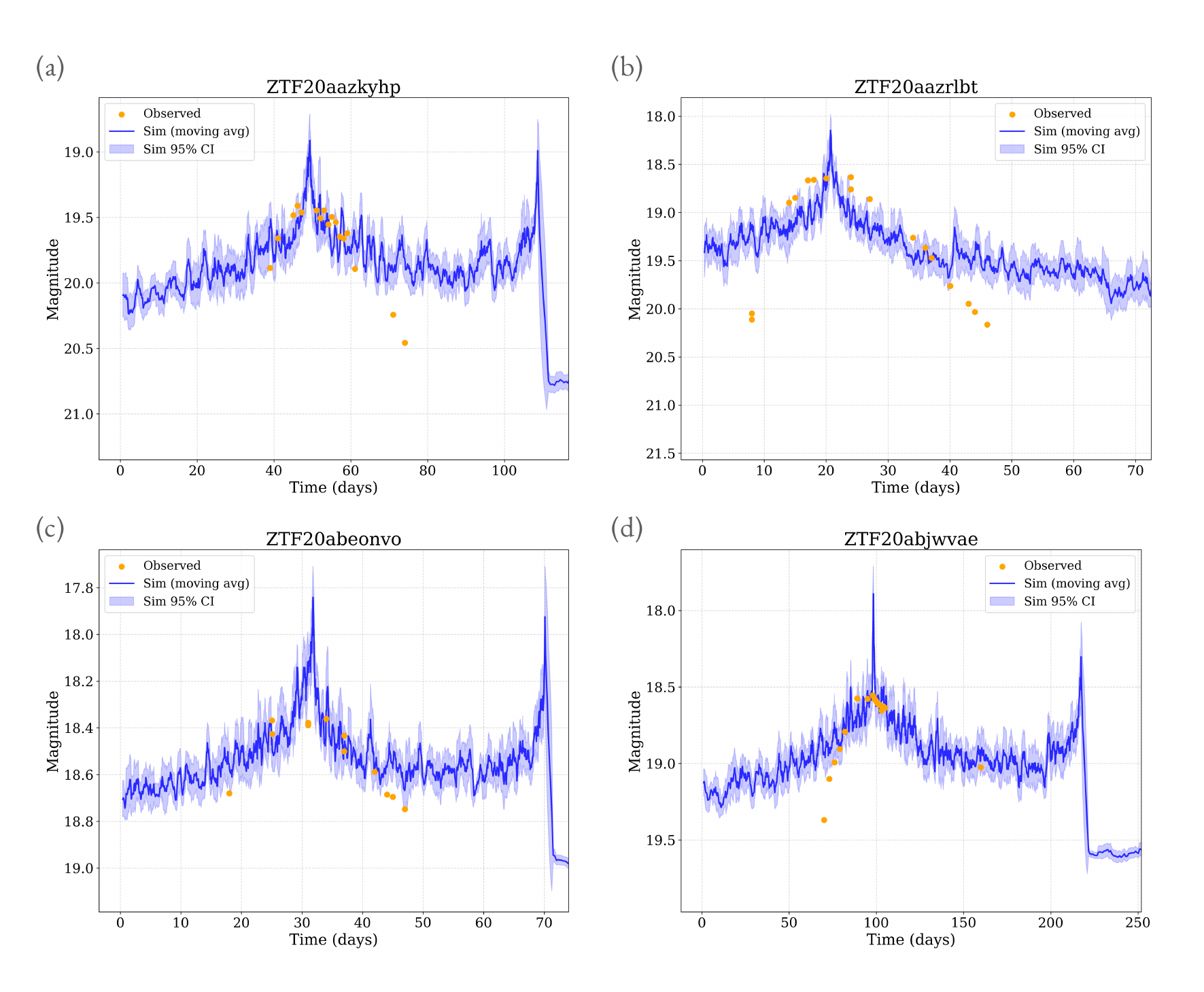}
  }
      \caption{Light curves for four transients: (a) ZTF20aazkyhp, (b) ZTF20aazrlbt, (c) ZTF20abeonvo, (d) ZTF20abjwvae. In each panel, the observed data from ZTF (orange points) are plotted as apparent magnitude versus time (in days, observer frame); a constant time offset has been applied to improve alignment with the simulation. The blue line shows the 10‑day moving average of the simulation, while the light blue shaded area denotes the corresponding 95\% confidence interval of the simulated data, calculated from the standard error of the simulation magnitudes over the 10-day windows and assuming a normal distribution. The physical parameters used in the simulations—as well as the corresponding known system properties—are summarized in Table \ref{tab:params}.}
         \label{fig:ch3_lcs_4}
   \end{figure}

\begin{figure}[htbp]
   \centering
   \makebox[\textwidth][c]{%
    \includegraphics[width=1\textwidth]{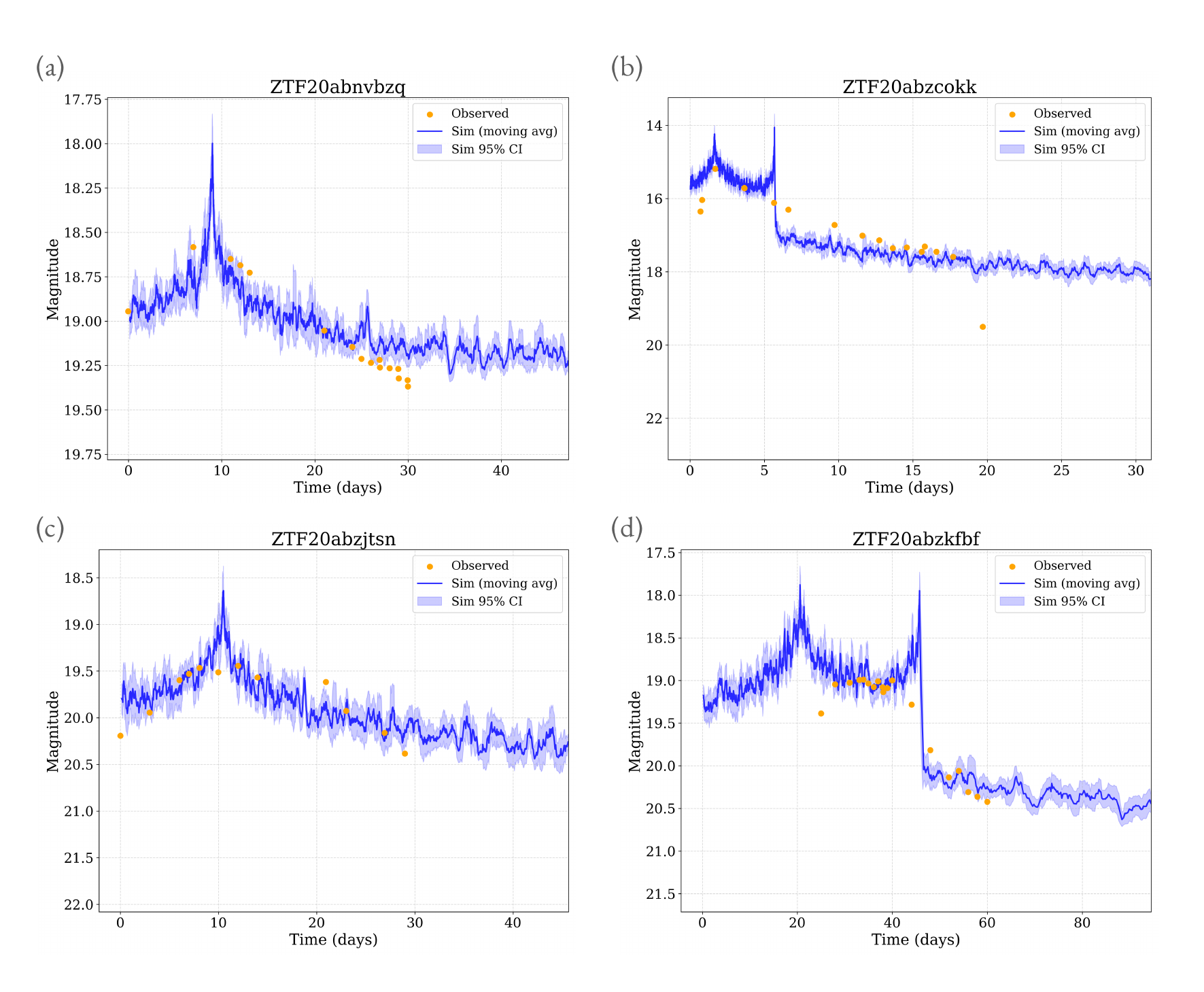}
  }
      \caption{Light curves for four transients: (a) ZTF20abnvbzq, (b) ZTF20abzcokk, (c) ZTF20abzjtsn, (d) ZTF20abzkfbf. In each panel, the observed data from ZTF (orange points) are plotted as apparent magnitude versus time (in days, observer frame); a constant time offset has been applied to improve alignment with the simulation. The blue line shows the 10‑day moving average of the simulation, while the light blue shaded area denotes the corresponding 95\% confidence interval of the simulated data, calculated from the standard error of the simulation magnitudes over the 10-day windows and assuming a normal distribution. The physical parameters used in the simulations—as well as the corresponding known system properties—are summarized in Table \ref{tab:params}.}
         \label{fig:ch3_lcs_5}
   \end{figure}

\begin{figure}[htbp]
   \centering
   \makebox[\textwidth][c]{%
    \includegraphics[width=1\textwidth]{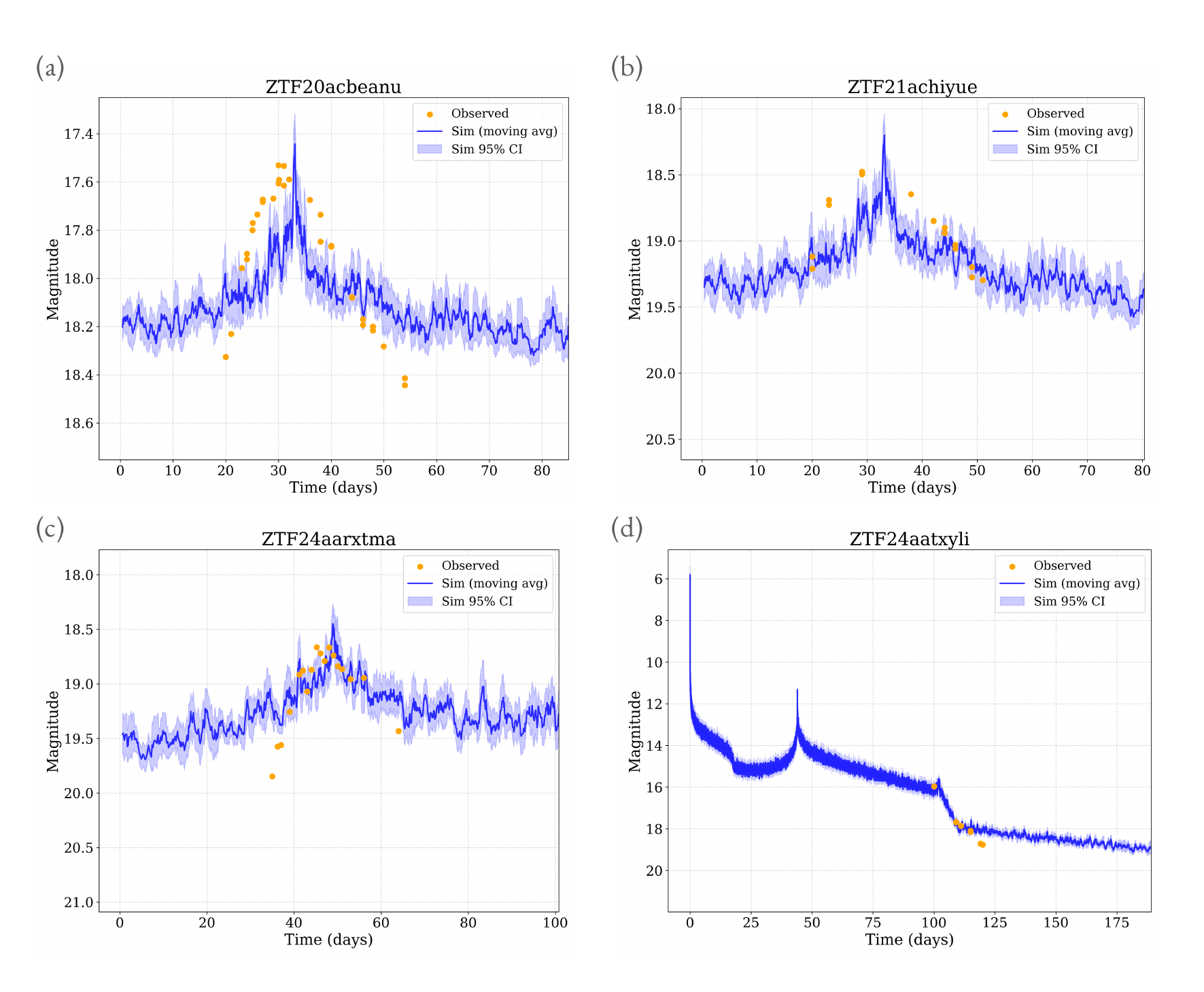}
  }
      \caption{Light curves for four transients: (a) ZTF20acbeanu, (b) ZTF21achiyue, (c) ZTF24aarxtma, (d) ZTF24aatxyli. In each panel, the observed data from ZTF (orange points) are plotted as apparent magnitude versus time (in days, observer frame); a constant time offset has been applied to improve alignment with the simulation. The blue line shows the 10‑day moving average of the simulation, while the light blue shaded area denotes the corresponding 95\% confidence interval of the simulated data, calculated from the standard error of the simulation magnitudes over the 10-day windows and assuming a normal distribution. The physical parameters used in the simulations—as well as the corresponding known system properties—are summarized in Table \ref{tab:params}.}
         \label{fig:ch3_lcs_6}
   \end{figure}

\begin{figure}[htbp]
   \centering
   \makebox[\textwidth][c]{%
    \includegraphics[width=1\textwidth]{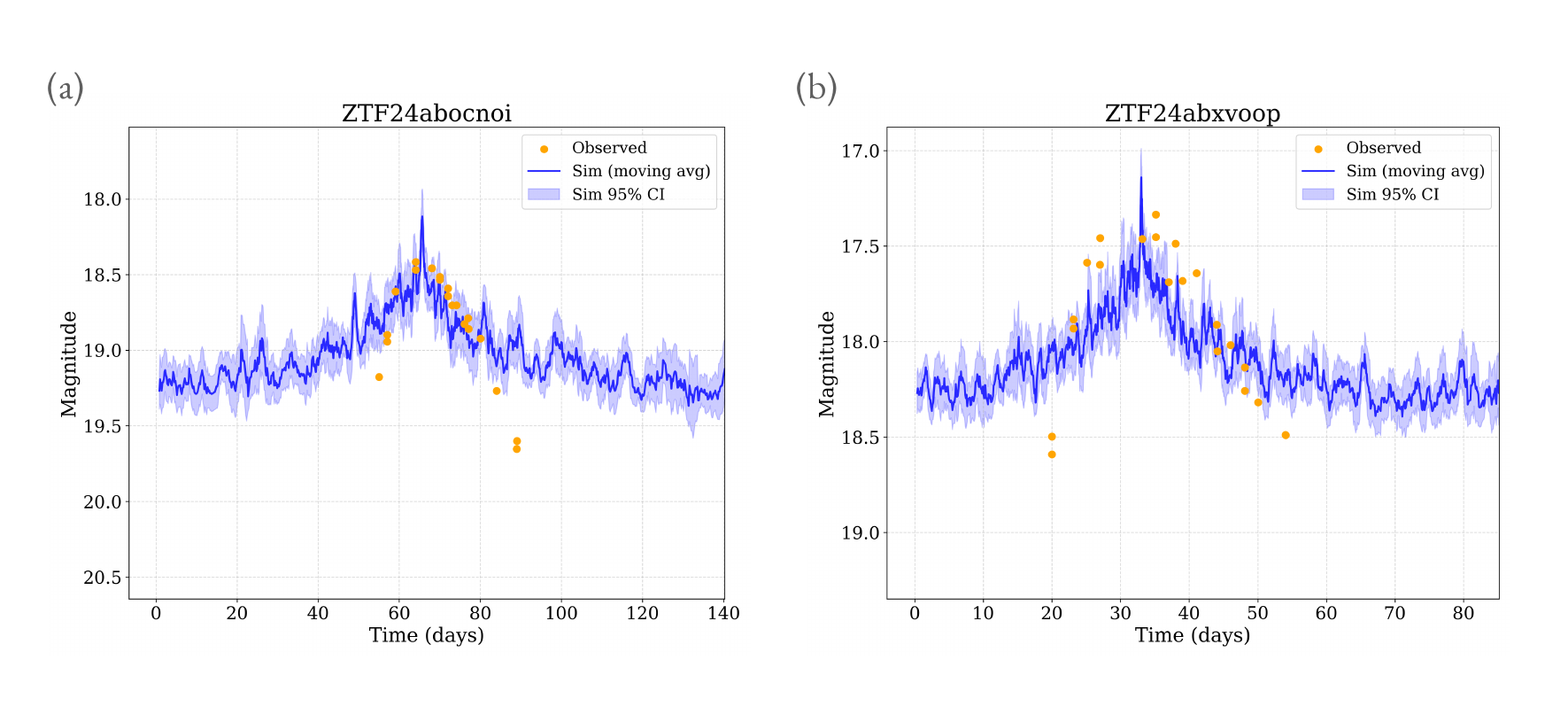}
  }
      \caption{Light curves for two transients: (a) ZTF24abocnoi, (b) ZTF24abxvoop. In each panel, the observed data from ZTF (orange points) are plotted as apparent magnitude versus time (in days, observer frame); a constant time offset has been applied to improve alignment with the simulation. The blue line shows the 10‑day moving average of the simulation, while the light blue shaded area denotes the corresponding 95\% confidence interval of the simulated data, calculated from the standard error of the simulation magnitudes over the 10-day windows and assuming a normal distribution. The physical parameters used in the simulations—as well as the corresponding known system properties—are summarized in Table \ref{tab:params}.}
         \label{fig:ch3_lcs_7}
   \end{figure}

In \Cref{fig:ch3_lcs_1,fig:ch3_lcs_2,fig:ch3_lcs_3,fig:ch3_lcs_4,fig:ch3_lcs_5,fig:ch3_lcs_6,fig:ch3_lcs_7}
, we display 26 ZTF light curves, each accompanied by a simulation light curve chosen via the method described in Section \ref{method}. The observed data points, in orange, show apparent magnitude as a function of time in the observer frame, starting at day 0 for the simulation. The solid blue line depicts the 10‐day moving average of the best‐matching simulation. The light blue shaded region indicates the 95\% confidence interval (CI) for this moving average, derived from the standard deviation of the simulation magnitudes in each 10‐day window. 

We find that many light curves observed by ZTF exhibit features present throughout the results of our simulations of stellar collisions, such as sharp rises and falls. Visually, it appears that most results do not have a late-time plateau behavior that we expect from many of our simulations; there are some exceptions, such as ZTF20abjwvae, ZTF20abnvbzq, and ZTF24aarxtma. Across most of our results, we can see that having access to higher-quality, longer-duration data (which we can expect with the start of LSST) will both help rule out and discover stellar collision candidates. 

We list various parameters of interest in Table \ref{tab:params}. For each transient we have plotted in \Cref{fig:ch3_lcs_1,fig:ch3_lcs_2,fig:ch3_lcs_3,fig:ch3_lcs_4,fig:ch3_lcs_5,fig:ch3_lcs_6,fig:ch3_lcs_7}
, we list its host galaxy's $g$ and $r$ magnitudes as provided in SDSS DR18~\cite{almeida2023}, if available. We list the transient's redshift $z$ as provided by Sherlock, if available. When not available, we default to a modest redshift value of $z=0.10$. We need to assume some value for calculation purposes, and this value is chosen based on other samples of ZTF-discovered transients~\cite{yao2019,bellm2019}, which find that ZTF-discovered transients such as SNe and TDEs are generally discovered at low redshifts ($z\lesssim0.2$). We list three different SMBH mass $M_{\bullet}$ values. The first two are calculated from photometric data, with one derived from a sample of local AGNs and the other assuming an elliptical/classical bulge, for largely quiescent SMBHs. The full detail of this calculation is provided in Appendix \ref{app:M_SMBH}. The third value of $M_{\bullet}$ listed in the table is the one used in the best-fit simulation. Similar values of the best-fit simulation's $M_{\bullet}$ value compared to either of the two photometry-derived values would be promising in terms of validating the robustness of our simulations. Finally, we provide the mass of the ejecta $M_{\mathrm{ej}}$ used in the simulation, as well as the constant offset applied when aligning the observed data with the best-fit simulation. 

We note that for sun-like stars, the tidal-disruption radius is smaller than the event horizon when the black hole mass exceeds roughly $10^8\,M_{\odot}$~\cite{dorazio2019}, beyond with a TDE cannot be observed. For maximally spinning black holes, TDEs can be observed for sun-like stars around SMBHs with masses up to approximately $7\times10^8\,M_{\odot}$~\cite{kesden2012}. As such, some of the largest candidates of interest we have identified cannot be associated with TDEs.

Our results demonstrate that a fully systematic pipeline can be used to identify candidate stellar collision events from a large volume stream of observed events by matching observed ZTF light curves to a large, pre-computed simulation bank using an approximate nearest neighbor algorithm. Furthermore, the method we have utilized incorporates multiple key physical parameters, such as supermassive black hole mass, ejecta mass, temporal offsets, and more, which both offers flexibility and provides the opportunity to test our model against known physical information, allowing us to further fine-tune our models and simulations as needed. This systematic approach minimizes the need for manual intervention, makes it less likely that we will only study exceptional or particularly unusual events, and enables robust parameter estimation even in the presence of complex light curve behavior.

Looking ahead, the forthcoming data stream from next-generation surveys like LSST will necessitate fully automated, scalable search methodologies. As LSST is expected to generate data at rates an order of magnitude greater than current surveys, the manual vetting of transient events will become impractical. Our pipeline, which leverages efficient data management techniques and advanced machine learning methods, is well positioned to meet this challenge. Furthermore, it is easy to both adapt and modify as needed and in response to advances in methodology. Moreover, the increased cadence and improved quality of LSST observations will not only enhance the reliability of candidate selection but also enable further refinement of our simulations and models. With higher-quality data, the discrepancies between simulated and observed light curves can be minimized, thereby advancing our understanding of the underlying physics governing stellar collisions in galactic nuclei.

\section{Discussion \& Future Work}\label{discussion}

In summary, we have developed and demonstrated the usability of a fully systematic pipeline for identifying candidate stellar collision events among transient light curves. Our method begins by generating an extensive bank of simulated light curves, each generated starting with a combination of physical parameters including supermassive black hole mass, initial radial position, and velocity orientation. Observed data from the Zwicky Transient Facility are filtered and then matched to the simulation bank using an approximate nearest neighbor algorithm implemented via the \texttt{annoy} library. This approach circumvents the prohibitive computational cost associated with traditional methods like MCMC, enabling rapid, automated matching between observations and simulations. The pipeline is able to successfully identify transients whose light curves are well-matched by our simulated models. Key features such as sharp rises, declines, and the overall morphology of the light curves are reproduced in the best-fit simulations. We additionally incorporate free parameters to adjust the temporal offset and luminosity scaling to both allow for better fitting and account for the uncertainty in the onset of the observed transient and the mass of the ejecta, respectively.

Looking to the future, the success of our systematic pipeline is particularly encouraging in light of the impending data influx from next-generation surveys such as the Vera C. Rubin Observatory's LSST. With LSST expected to produce data volumes that dwarf current surveys by an order of magnitude, automated and scalable search methodologies like ours will be essential. The capability to rapidly and reliably sift through vast datasets to identify promising transient candidates will not only streamline the discovery process but also facilitate timely follow-up observations and detailed analyses.

Furthermore, the advent of higher-cadence and higher-quality data will provide an opportunity to refine our simulation models and parameter estimation techniques. Improved observational data will enable more precise tuning of our simulations, potentially revealing subtle trends and correlations that were previously obscured by lower signal-to-noise ratios. This iterative process of model refinement and validation is expected to yield deeper insights into the underlying physics of stellar collisions in galactic nuclei, thereby enhancing our overall understanding of these rare and energetic phenomena.

The method we have described in this work is easy to both implement and adapt. The library we use, \texttt{annoy}, has Python bindings, making it very user-friendly, over a C++ implementation, which allows for better performance. In future work, it would be wise to explore alternatives, especially as machine learning methods are constantly advancing. Alternative methods that could be explored include traditional data structures such as $k$-d-trees or ball trees for exact searches in lower dimensions, as well as probabilistic approaches like Locality-Sensitive Hashing that can handle higher-dimensional data efficiently. More advanced techniques such as Hierarchical Navigable Small World (HNSW) graphs~\cite{malkov2018} and selecting between randomized $k$-d tree algorithms and hierarchical $k$-means tree algorithms, as implemented in FLANN~\cite{muja2009} also show promise in balancing speed and accuracy, and they warrant further investigation as machine learning and data processing methods continue to evolve.

\onecolumn
\renewcommand{\arraystretch}{1.2} 
\begin{table}[h]
    \centering
    \caption{The set of criteria used to identify high-quality TDE and stellar collision candidates. In this context, Sherlock is the software used by Lasair that cross-matches aggregated transient detections against archival catalogs, providing context-based classifications for each detected object~\cite{smith2020}. The object can be classified as one of [NULL, AGN, BS, CV, NT, ORPHAN, SN, UNCLEAR, VS], with full descriptions of each classification provided in the Lasair documentation.}
    \label{tab:lasair_criteria}
    \begin{tabularx}{\textwidth}{>{\raggedright\arraybackslash}m{0.40\textwidth} >{\raggedright\arraybackslash}m{0.40\textwidth}}
        \toprule
        \multicolumn{1}{c}{\textbf{Condition}} & \multicolumn{1}{c}{\textbf{Interpretation}} \\
        \midrule
        \texttt{sherlock\_classifications .separationArcsec < 1.0} 
            & Object is within 1 arcsec of Sherlock designated host \\
        \midrule
        \texttt{objects.ncand > 20} 
            & Number of detections is more than 20 \\
        \midrule
        \texttt{objects.ncandgp > 10} 
            & More than 10 detections of positive flux (and "good" quality) \\
        \midrule
        \texttt{sherlock\_classifications .classification in ('NT','SN')}
            & Sherlock thinks the object is a nuclear transient or SNe \\
        \midrule
        \texttt{objects.maggmean - objects.magrmean < 0.05 OR objects.gmag - objects.rmag < 0.05}
            & Mean color is blue or latest color is blue \\
        \midrule
        \texttt{objects.rmag < 20.0 OR objects.gmag < 20.0}
            & Object is brighter than 20 in $r$ and $g$ \\
        \midrule
        \texttt{objects.glatmean > 10 OR objects.glatmean < - 10}
            & Object is not within $\pm10$ degrees of galactic latitude \\
        \midrule
        \texttt{objects.sgmag1 - objects.srmag1 > 0}
            & Possible PS1 host mag is red \\
        \bottomrule
    \end{tabularx}
\end{table}

\onecolumn
\clearpage
   \clearpage
   \clearpage
   \clearpage
   \clearpage
   \clearpage
   \clearpage
   
   \begin{table}[h]
    \centering
    \caption{Summary of properties for the selected transients. Columns list the transient identifier, the $g$ and $r$ magnitudes of the host galaxy from SDSS DR18 (when available), the redshift (as provided by Sherlock, or a default value of $0.10^{\dag}$ if unavailable), and the object classification according to Sherlock ($^{\ddag}$NT = nuclear transient, SN = supernova). Three values of logarithmic SMBH masses are given; they are derived from the photometry, both for local AGNs and under the assumption of an elliptical/classical bulge (from largely quiescent SMBHs), as well as the logarithmic SMBH mass used in the simulation. The table further shows the ejecta mass adopted in the simulation and the constant time offset (in days) applied to the observed data.}
    \label{tab:params}
    \renewcommand{\arraystretch}{1.2}
    \makebox[\linewidth][c]{%
    \begin{tabular}{l c c c c c c c c c}
        \toprule
        \midrule
        \multicolumn{1}{c}{Transient} & $g$ & $r$ & Redshift & Class$\,^{\ddag}$ & 
        $\log \frac{M_{\bullet,\mathrm{AGN}}}{M_{\odot}}$ & 
        $\log \frac{M_{\bullet,\mathrm{ell}}}{M_{\odot}}$ & 
        $\log \frac{M_{\bullet,\mathrm{sim}}}{M_{\odot}}$ & 
        $M_{\mathrm{ej}}\,[M_{\odot}]$ & Offset [days] \\
        \midrule
	ZTF18aarlukc & 17.16 & 16.33 & $0.06$ & NT & 7.27 & 8.71 & 9.00 & 0.02 & 3 \\
	ZTF19abbwbrw & 21.28 & 20.70 & $0.10^{\dag}$ & SN & 5.64 & 6.53 & 8.00 & 0.01 & -8 \\
	ZTF19abcvrpm & 22.06 & 21.60 & $0.10^{\dag}$ & SN & 5.12 & 5.84 & 7.70 & 0.01 & -30 \\
	ZTF19acchink & 20.48 & 19.63 & $0.09$ & SN & 6.30 & 7.41 & 7.70 & 0.01 & -28 \\
	ZTF19acgonwy & N/A & N/A & $0.10^{\dag}$ & SN & N/A & N/A & 8.00 & 0.03 & 0 \\
	ZTF20aabqiqd & 20.00 & 19.46 & $0.10^{\dag}$ & SN & 6.11 & 7.17 & 8.00 & 0.01 & -3 \\
	ZTF20aagnrdd & N/A & N/A & $0.10^{\dag}$ & SN & N/A & N/A & 8.00 & 0.02 & -3 \\
	ZTF20aajcbhr & 19.04 & 18.65 & $0.10^{\dag}$ & NT & 6.28 & 7.39 & 7.70 & 0.01 & 2 \\
	ZTF20aakkluy & N/A & N/A & $0.10^{\dag}$ & SN & N/A & N/A & 9.00 & 0.45 & 0 \\
	ZTF20aavrhnd & N/A & N/A & $0.10^{\dag}$ & SN & N/A & N/A & 7.70 & 0.01 & -10 \\
	ZTF20aavsjuv & N/A & N/A & $0.10^{\dag}$ & SN & N/A & N/A & 8.00 & 0.02 & 0 \\
	ZTF20aawoojh & 15.61 & 15.02 & $0.10^{\dag}$ & NT & 8.03 & 9.73 & 8.00 & 0.01 & -5 \\
	ZTF20aazkyhp & 20.93 & 20.28 & $0.10^{\dag}$ & SN & 5.89 & 6.88 & 7.70 & 0.01 & -11 \\
	ZTF20aazrlbt & 20.00 & 19.69 & $0.10^{\dag}$ & SN & 5.75 & 6.68 & 8.00 & 0.01 & 8 \\
	ZTF20abeonvo & 18.04 & 17.28 & $0.10^{\dag}$ & NT & 7.28 & 8.72 & 8.00 & 0.01 & -7 \\
	ZTF20abjwvae & 19.74 & 18.97 & $0.10^{\dag}$ & SN & 6.58 & 7.79 & 8.00 & 0.03 & 20 \\
	ZTF20abnvbzq & N/A & N/A & $0.10^{\dag}$ & SN & N/A & N/A & 8.70 & 0.00 & -3 \\
	ZTF20abzcokk & 22.39 & 22.50 & $0.10^{\dag}$ & SN & 4.09 & 4.46 & 6.70 & 0.03 & 0 \\
	ZTF20abzjtsn & 21.98 & 21.44 & $0.10^{\dag}$ & SN & 5.28 & 6.06 & 7.70 & 0.00 & 0 \\
	ZTF20abzkfbf & 21.32 & 20.97 & $0.10^{\dag}$ & SN & 5.26 & 6.03 & 5.70 & 0.01 & 0 \\
	ZTF20acbeanu & N/A & N/A & $0.10^{\dag}$ & NT & N/A & N/A & 8.00 & 0.02 & -5 \\
	ZTF21achiyue & 19.24 & 18.71 & $0.10^{\dag}$ & SN & 6.42 & 7.57 & 8.00 & 0.01 & -5 \\
	ZTF24aarxtma & N/A & N/A & $0.10^{\dag}$ & SN & N/A & N/A & 7.70 & 0.02 & 10 \\
	ZTF24aatxyli & 22.52 & 22.15 & $0.10^{\dag}$ & SN & 4.79 & 5.40 & 5.70 & 0.02 & 0 \\
	ZTF24abocnoi & 20.03 & 19.46 & $0.10^{\dag}$ & SN & 6.15 & 7.21 & 6.70 & 0.03 & 5 \\
	ZTF24abxvoop & 18.71 & 17.89 & $0.10^{\dag}$ & SN & 7.09 & 8.47 & 8.00 & 0.03 & -5 \\
        \midrule
        \bottomrule
    \end{tabular}
    }
\end{table}

\begin{acknowledgements}
      B.X.H. and A.L. acknowledge support from the Black Hole Initiative, which is supported by the John Templeton Foundation and the Gordon and Betty Moore Foundation. B.X.H. acknowledges support from the Department of Defense National Defense Science and Engineering Graduate Fellowship. \end{acknowledgements}

%
   \bibliographystyle{aa} 
   \bibliography{references} 

\begin{thebibliography}{47}
\expandafter\ifx\csname natexlab\endcsname\relax\def\natexlab#1{#1}\fi

\bibitem[{{Almeida} {et~al.}(2023){Almeida}, {Anderson}, {Argudo-Fern{\'a}ndez}, {Badenes}, {Barger}, {Barrera-Ballesteros}, {Bender}, {Benitez}, {Besser}, {Bird}, {Bizyaev}, {Blanton}, {Bochanski}, {Bovy}, {Brandt}, {Brownstein}, {Buchner}, {Bulbul}, {Burchett}, {Cano D{\'\i}az}, {Carlberg}, {Casey}, {Chandra}, {Cherinka}, {Chiappini}, {Coker}, {Comparat}, {Conroy}, {Contardo}, {Cortes}, {Covey}, {Crane}, {Cunha}, {Dabbieri}, {Davidson}, {Davis}, {de Andrade Queiroz}, {De Lee}, {M{\'e}ndez Delgado}, {Demasi}, {Di Mille}, {Donor}, {Dow}, {Dwelly}, {Eracleous}, {Eriksen}, {Fan}, {Farr}, {Frederick}, {Fries}, {Frinchaboy}, {G{\"a}nsicke}, {Ge}, {Gonz{\'a}lez {\'A}vila}, {Grabowski}, {Grier}, {Guiglion}, {Gupta}, {Hall}, {Hawkins}, {Hayes}, {Hermes}, {Hern{\'a}ndez-Garc{\'\i}a}, {Hogg}, {Holtzman}, {Ibarra-Medel}, {Ji}, {Jofre}, {Johnson}, {Jones}, {Kinemuchi}, {Kluge}, {Koekemoer}, {Kollmeier}, {Kounkel}, {Krishnarao}, {Krumpe}, {Lacerna}, {Lago}, {Laporte}, {Liu}, {Liu}, {Liu}, {Lopes}, {Macktoobian},
  {Majewski}, {Malanushenko}, {Maoz}, {Masseron}, {Masters}, {Matijevic}, {McBride}, {Medan}, {Merloni}, {Morrison}, {Myers}, {M{\'e}sz{\'a}ros}, {Negrete}, {Nidever}, {Nitschelm}, {Oravetz}, {Oravetz}, {Pan}, {Peng}, {Pinsonneault}, {Pogge}, {Qiu}, {Ramirez}, {Rix}, {Fern{\'a}ndez Rosso}, {Runnoe}, {Salvato}, {Sanchez}, {Santana}, {Saydjari}, {Sayres}, {Schlaufman}, {Schneider}, {Schwope}, {Serna}, {Shen}, {Sobeck}, {Song}, {Souto}, {Spoo}, {Stassun}, {Steinmetz}, {Straumit}, {Stringfellow}, {S{\'a}nchez-Gallego}, {Taghizadeh-Popp}, {Tayar}, {Thakar}, {Tissera}, {Tkachenko}, {Hernandez Toledo}, {Trakhtenbrot}, {Fern{\'a}ndez-Trincado}, {Troup}, {Trump}, {Tuttle}, {Ulloa}, {Vazquez-Mata}, {Vera Alfaro}, {Villanova}, {Wachter}, {Weijmans}, {Wheeler}, {Wilson}, {Wojno}, {Wolf}, {Xue}, {Ybarra}, {Zari}, \& {Zasowski}}]{almeida2023}
{Almeida}, A., {Anderson}, S.~F., {Argudo-Fern{\'a}ndez}, M., {et~al.} 2023, \apjs, 267, 44

\bibitem[{{Arnett}(1996)}]{arnett1996}
{Arnett}, D. 1996, {Supernovae and Nucleosynthesis: An Investigation of the History of Matter from the Big Bang to the Present}

\bibitem[{{Astropy Collaboration} {et~al.}(2022){Astropy Collaboration}, {Price-Whelan}, {Lim}, {Earl}, {Starkman}, {Bradley}, {Shupe}, {Patil}, {Corrales}, {Brasseur}, {N{\"o}the}, {Donath}, {Tollerud}, {Morris}, {Ginsburg}, {Vaher}, {Weaver}, {Tocknell}, {Jamieson}, {van Kerkwijk}, {Robitaille}, {Merry}, {Bachetti}, {G{\"u}nther}, {Aldcroft}, {Alvarado-Montes}, {Archibald}, {B{\'o}di}, {Bapat}, {Barentsen}, {Baz{\'a}n}, {Biswas}, {Boquien}, {Burke}, {Cara}, {Cara}, {Conroy}, {Conseil}, {Craig}, {Cross}, {Cruz}, {D'Eugenio}, {Dencheva}, {Devillepoix}, {Dietrich}, {Eigenbrot}, {Erben}, {Ferreira}, {Foreman-Mackey}, {Fox}, {Freij}, {Garg}, {Geda}, {Glattly}, {Gondhalekar}, {Gordon}, {Grant}, {Greenfield}, {Groener}, {Guest}, {Gurovich}, {Handberg}, {Hart}, {Hatfield-Dodds}, {Homeier}, {Hosseinzadeh}, {Jenness}, {Jones}, {Joseph}, {Kalmbach}, {Karamehmetoglu}, {Ka{\l}uszy{\'n}ski}, {Kelley}, {Kern}, {Kerzendorf}, {Koch}, {Kulumani}, {Lee}, {Ly}, {Ma}, {MacBride}, {Maljaars}, {Muna}, {Murphy}, {Norman},
  {O'Steen}, {Oman}, {Pacifici}, {Pascual}, {Pascual-Granado}, {Patil}, {Perren}, {Pickering}, {Rastogi}, {Roulston}, {Ryan}, {Rykoff}, {Sabater}, {Sakurikar}, {Salgado}, {Sanghi}, {Saunders}, {Savchenko}, {Schwardt}, {Seifert-Eckert}, {Shih}, {Jain}, {Shukla}, {Sick}, {Simpson}, {Singanamalla}, {Singer}, {Singhal}, {Sinha}, {Sip{\H{o}}cz}, {Spitler}, {Stansby}, {Streicher}, {{\v{S}}umak}, {Swinbank}, {Taranu}, {Tewary}, {Tremblay}, {de Val-Borro}, {Van Kooten}, {Vasovi{\'c}}, {Verma}, {de Miranda Cardoso}, {Williams}, {Wilson}, {Winkel}, {Wood-Vasey}, {Xue}, {Yoachim}, {Zhang}, {Zonca}, \& {Astropy Project Contributors}}]{astropy2022}
{Astropy Collaboration}, {Price-Whelan}, A.~M., {Lim}, P.~L., {et~al.} 2022, \apj, 935, 167

\bibitem[{Ball \& Brunner(2010)}]{ball2010}
Ball, N.~M. \& Brunner, R.~J. 2010, International Journal of Modern Physics D, 19, 1049–1106

\bibitem[{Baron(2019)}]{baron2019}
Baron, D. 2019, Machine Learning in Astronomy: a practical overview

\bibitem[{{Bell} {et~al.}(2003){Bell}, {McIntosh}, {Katz}, \& {Weinberg}}]{bell2003}
{Bell}, E.~F., {McIntosh}, D.~H., {Katz}, N., \& {Weinberg}, M.~D. 2003, \apjs, 149, 289

\bibitem[{{Bellm} {et~al.}(2019){Bellm}, {Kulkarni}, {Graham}, {Dekany}, {Smith}, {Riddle}, {Masci}, {Helou}, {Prince}, {Adams}, {Barbarino}, {Barlow}, {Bauer}, {Beck}, {Belicki}, {Biswas}, {Blagorodnova}, {Bodewits}, {Bolin}, {Brinnel}, {Brooke}, {Bue}, {Bulla}, {Burruss}, {Cenko}, {Chang}, {Connolly}, {Coughlin}, {Cromer}, {Cunningham}, {De}, {Delacroix}, {Desai}, {Duev}, {Eadie}, {Farnham}, {Feeney}, {Feindt}, {Flynn}, {Franckowiak}, {Frederick}, {Fremling}, {Gal-Yam}, {Gezari}, {Giomi}, {Goldstein}, {Golkhou}, {Goobar}, {Groom}, {Hacopians}, {Hale}, {Henning}, {Ho}, {Hover}, {Howell}, {Hung}, {Huppenkothen}, {Imel}, {Ip}, {Ivezi{\'c}}, {Jackson}, {Jones}, {Juric}, {Kasliwal}, {Kaspi}, {Kaye}, {Kelley}, {Kowalski}, {Kramer}, {Kupfer}, {Landry}, {Laher}, {Lee}, {Lin}, {Lin}, {Lunnan}, {Giomi}, {Mahabal}, {Mao}, {Miller}, {Monkewitz}, {Murphy}, {Ngeow}, {Nordin}, {Nugent}, {Ofek}, {Patterson}, {Penprase}, {Porter}, {Rauch}, {Rebbapragada}, {Reiley}, {Rigault}, {Rodriguez}, {van Roestel}, {Rusholme}, {van
  Santen}, {Schulze}, {Shupe}, {Singer}, {Soumagnac}, {Stein}, {Surace}, {Sollerman}, {Szkody}, {Taddia}, {Terek}, {Van Sistine}, {van Velzen}, {Vestrand}, {Walters}, {Ward}, {Ye}, {Yu}, {Yan}, \& {Zolkower}}]{bellm2019}
{Bellm}, E.~C., {Kulkarni}, S.~R., {Graham}, M.~J., {et~al.} 2019, \pasp, 131, 018002

\bibitem[{Bentley(1975)}]{bentley1975}
Bentley, J.~L. 1975, Commun. ACM, 18, 509–517

\bibitem[{Bernhardsson(2018)}]{annoy}
Bernhardsson, E. 2018, Annoy: Approximate Nearest Neighbors in C++/Python, python package version 1.13.0

\bibitem[{{Childress} {et~al.}(2017){Childress}, {Lidman}, {Davis}, {Tucker}, {Asorey}, {Yuan}, {Abbott}, {Abdalla}, {Allam}, {Annis}, {Banerji}, {Benoit-L{\'e}vy}, {Bernard}, {Bertin}, {Brooks}, {Buckley-Geer}, {Burke}, {Carnero Rosell}, {Carollo}, {Carrasco Kind}, {Carretero}, {Castander}, {Cunha}, {da Costa}, {D'Andrea}, {Doel}, {Eifler}, {Evrard}, {Flaugher}, {Foley}, {Fosalba}, {Frieman}, {Garc{\'\i}a-Bellido}, {Glazebrook}, {Goldstein}, {Gruen}, {Gruendl}, {Gschwend}, {Gupta}, {Gutierrez}, {Hinton}, {Hoormann}, {James}, {Kessler}, {Kim}, {King}, {Kovacs}, {Kuehn}, {Kuhlmann}, {Kuropatkin}, {Lagattuta}, {Lewis}, {Li}, {Lima}, {Lin}, {Macaulay}, {Maia}, {Marriner}, {March}, {Marshall}, {Martini}, {McMahon}, {Menanteau}, {Miquel}, {Moller}, {Morganson}, {Mould}, {Mudd}, {Muthukrishna}, {Nichol}, {Nord}, {Ogando}, {Ostrovski}, {Parkinson}, {Plazas}, {Reed}, {Reil}, {Romer}, {Rykoff}, {Sako}, {Sanchez}, {Scarpine}, {Schindler}, {Schubnell}, {Scolnic}, {Sevilla-Noarbe}, {Seymour}, {Sharp}, {Smith},
  {Soares-Santos}, {Sobreira}, {Sommer}, {Spinka}, {Suchyta}, {Sullivan}, {Swanson}, {Tarle}, {Uddin}, {Walker}, {Wester}, \& {Zhang}}]{childress2017}
{Childress}, M.~J., {Lidman}, C., {Davis}, T.~M., {et~al.} 2017, \mnras, 472, 273

\bibitem[{Cote {et~al.}(2006)Cote, Piatek, Ferrarese, Jordan, Merritt, Peng, Haşegan, Blakeslee, Mei, West, Milosavljević, \& Tonry}]{cote2006}
Cote, P., Piatek, S., Ferrarese, L., {et~al.} 2006, The Astrophysical Journal Supplement Series, 165, 57–94

\bibitem[{Dieleman {et~al.}(2015)Dieleman, Willett, \& Dambre}]{dieleman2015}
Dieleman, S., Willett, K.~W., \& Dambre, J. 2015, Monthly Notices of the Royal Astronomical Society, 450, 1441–1459

\bibitem[{Domínguez~Sánchez {et~al.}(2018)Domínguez~Sánchez, Huertas-Company, Bernardi, Tuccillo, \& Fischer}]{dominguezsanchez2018}
Domínguez~Sánchez, H., Huertas-Company, M., Bernardi, M., Tuccillo, D., \& Fischer, J.~L. 2018, Monthly Notices of the Royal Astronomical Society, 476, 3661–3676

\bibitem[{{D'Orazio} {et~al.}(2019){D'Orazio}, {Loeb}, \& {Guillochon}}]{dorazio2019}
{D'Orazio}, D.~J., {Loeb}, A., \& {Guillochon}, J. 2019, \mnras, 485, 4413

\bibitem[{du~Buisson {et~al.}(2015)du~Buisson, Sivanandam, Bassett, \& Smith}]{dubuisson2015}
du~Buisson, L., Sivanandam, N., Bassett, B.~A., \& Smith, M. 2015, Monthly Notices of the Royal Astronomical Society, 454, 2026–2038

\bibitem[{Fluke \& Jacobs(2019)}]{fluke2019}
Fluke, C.~J. \& Jacobs, C. 2019, WIREs Data Mining and Knowledge Discovery, 10

\bibitem[{{Foreman-Mackey} {et~al.}(2013){Foreman-Mackey}, {Hogg}, {Lang}, \& {Goodman}}]{foremanmackey2013}
{Foreman-Mackey}, D., {Hogg}, D.~W., {Lang}, D., \& {Goodman}, J. 2013, \pasp, 125, 306

\bibitem[{French {et~al.}(2016)French, Arcavi, \& Zabludoff}]{french2016}
French, K.~D., Arcavi, I., \& Zabludoff, A. 2016, The Astrophysical Journal Letters, 818, L21

\bibitem[{Förster {et~al.}(2021)Förster, Cabrera-Vives, Castillo-Navarrete, Estévez, Sánchez-Sáez, Arredondo, Bauer, Carrasco-Davis, Catelan, Elorrieta, Eyheramendy, Huijse, Pignata, Reyes, Reyes, Rodríguez-Mancini, Ruz-Mieres, Valenzuela, Álvarez Maldonado, Astorga, Borissova, Clocchiatti, De~Cicco, Donoso-Oliva, Hernández-García, Graham, Jordán, Kurtev, Mahabal, Maureira, Muñoz-Arancibia, Molina-Ferreiro, Moya, Palma, Pérez-Carrasco, Protopapas, Romero, Sabatini-Gacitua, Sánchez, Martín, Sepúlveda-Cobo, Vera, \& Vergara}]{forster2021}
Förster, F., Cabrera-Vives, G., Castillo-Navarrete, E., {et~al.} 2021, The Astronomical Journal, 161, 242

\bibitem[{Gabbard {et~al.}(2022)Gabbard, Messenger, Heng, Tonolini, \& Murray-Smith}]{gabbard2022}
Gabbard, H., Messenger, C., Heng, I.~S., Tonolini, F., \& Murray-Smith, R. 2022, Nature Physics, 18, 112

\bibitem[{Gal \& Ghahramani(2016)}]{gal2016}
Gal, Y. \& Ghahramani, Z. 2016, in Proceedings of Machine Learning Research, Vol.~48, Proceedings of The 33rd International Conference on Machine Learning, ed. M.~F. Balcan \& K.~Q. Weinberger (New York, New York, USA: PMLR), 1050--1059

\bibitem[{{Gezari}(2021)}]{gezari2021}
{Gezari}, S. 2021, \araa, 59, 21

\bibitem[{{Gieseke} {et~al.}(2017){Gieseke}, {Bloemen}, {van den Bogaard}, {Heskes}, {Kindler}, {Scalzo}, {Ribeiro}, {van Roestel}, {Groot}, {Yuan}, {M{\"o}ller}, \& {Tucker}}]{giuseke2017}
{Gieseke}, F., {Bloemen}, S., {van den Bogaard}, C., {et~al.} 2017, \mnras, 472, 3101

\bibitem[{Hastings(1970)}]{hastings1970}
Hastings, W.~K. 1970, Biometrika, 57, 97

\bibitem[{{Hernquist}(1990)}]{hernquist1990}
{Hernquist}, L. 1990, \apj, 356, 359

\bibitem[{{Hu} \& {Loeb}(2024{\natexlab{a}})}]{hu2024b}
{Hu}, B.~X. \& {Loeb}, A. 2024{\natexlab{a}}, \aap, 690, A130

\bibitem[{{Hu} \& {Loeb}(2024{\natexlab{b}})}]{hu2024a}
{Hu}, B.~X. \& {Loeb}, A. 2024{\natexlab{b}}, \aap, 689, A23

\bibitem[{Kendall \& Gal(2017)}]{kendall2017}
Kendall, A. \& Gal, Y. 2017, What Uncertainties Do We Need in Bayesian Deep Learning for Computer Vision?

\bibitem[{{Kesden}(2012)}]{kesden2012}
{Kesden}, M. 2012, \prd, 85, 024037

\bibitem[{Lewis \& Bridle(2002)}]{lewis2002}
Lewis, A. \& Bridle, S. 2002, Physical Review D, 66

\bibitem[{Malkov \& Yashunin(2018)}]{malkov2018}
Malkov, Y.~A. \& Yashunin, D.~A. 2018, Efficient and robust approximate nearest neighbor search using Hierarchical Navigable Small World graphs

\bibitem[{Muja \& Lowe(2009)}]{muja2009}
Muja, M. \& Lowe, D.~G. 2009, VISAPP (1), 2, 2

\bibitem[{Muthukrishna {et~al.}(2019)Muthukrishna, Narayan, Mandel, Biswas, \& Hložek}]{muthukrishna2019}
Muthukrishna, D., Narayan, G., Mandel, K.~S., Biswas, R., \& Hložek, R. 2019, Publications of the Astronomical Society of the Pacific, 131, 118002

\bibitem[{Narayan {et~al.}(2019)Narayan, Johnson, \& Gammie}]{narayan2019}
Narayan, R., Johnson, M.~D., \& Gammie, C.~F. 2019, The Astrophysical Journal Letters, 885, L33

\bibitem[{{Nordin} {et~al.}(2019){Nordin}, {Brinnel}, {van Santen}, {Bulla}, {Feindt}, {Franckowiak}, {Fremling}, {Gal-Yam}, {Giomi}, {Kowalski}, {Mahabal}, {Miranda}, {Rauch}, {Reusch}, {Rigault}, {Schulze}, {Sollerman}, {Stein}, {Yaron}, {van Velzen}, \& {Ward}}]{nordin2019}
{Nordin}, J., {Brinnel}, V., {van Santen}, J., {et~al.} 2019, \aap, 631, A147

\bibitem[{Portillo {et~al.}(2020)Portillo, Parejko, Vergara, \& Connolly}]{portillo2020}
Portillo, S. K.~N., Parejko, J.~K., Vergara, J.~R., \& Connolly, A.~J. 2020, The Astronomical Journal, 160, 45

\bibitem[{{Rau} {et~al.}(2009){Rau}, {Kulkarni}, {Law}, {Bloom}, {Ciardi}, {Djorgovski}, {Fox}, {Gal-Yam}, {Grillmair}, {Kasliwal}, {Nugent}, {Ofek}, {Quimby}, {Reach}, {Shara}, {Bildsten}, {Cenko}, {Drake}, {Filippenko}, {Helfand}, {Helou}, {Howell}, {Poznanski}, \& {Sullivan}}]{rau2009}
{Rau}, A., {Kulkarni}, S.~R., {Law}, N.~M., {et~al.} 2009, \pasp, 121, 1334

\bibitem[{{Reines} \& {Volonteri}(2015)}]{reines2015}
{Reines}, A.~E. \& {Volonteri}, M. 2015, \apj, 813, 82

\bibitem[{{Shakura} \& {Sunyaev}(1973)}]{shakura1973}
{Shakura}, N.~I. \& {Sunyaev}, R.~A. 1973, \aap, 24, 337

\bibitem[{{Smith}(2019)}]{smith2019}
{Smith}, K. 2019, in The Extragalactic Explosive Universe: the New Era of Transient Surveys and Data-Driven Discovery, 51

\bibitem[{Smith {et~al.}(2020)Smith, Smartt, Young, Tonry, Denneau, Flewelling, Heinze, Weiland, Stalder, Rest, Stubbs, Anderson, Chen, Clark, Do, Förster, Fulton, Gillanders, McBrien, O’Neill, Srivastav, \& Wright}]{smith2020}
Smith, K.~W., Smartt, S.~J., Young, D.~R., {et~al.} 2020, Publications of the Astronomical Society of the Pacific, 132, 085002

\bibitem[{{Storey-Fisher} {et~al.}(2021){Storey-Fisher}, {Huertas-Company}, {Ramachandra}, {Lanusse}, {Leauthaud}, {Luo}, {Huang}, \& {Prochaska}}]{storeyfisher2021}
{Storey-Fisher}, K., {Huertas-Company}, M., {Ramachandra}, N., {et~al.} 2021, \mnras, 508, 2946

\bibitem[{Thorne {et~al.}(2021)Thorne, Knox, \& Prabhu}]{thorne2021}
Thorne, B., Knox, L., \& Prabhu, K. 2021, Monthly Notices of the Royal Astronomical Society, 504, 2603

\bibitem[{Tremaine {et~al.}(1994)Tremaine, Richstone, Byun, Dressler, Faber, Grillmair, Kormendy, \& Lauer}]{tremaine1994}
Tremaine, S., Richstone, D.~O., Byun, Y.-I., {et~al.} 1994, The Astronomical Journal, 107, 634

\bibitem[{Williams {et~al.}(2024)Williams, Francis, Lawrence, Sloan, Smartt, Smith, \& Young}]{williams2024}
Williams, R.~D., Francis, G.~P., Lawrence, A., {et~al.} 2024, Enabling Science from the Rubin Alert Stream with Lasair

\bibitem[{{Yao} {et~al.}(2019){Yao}, {Miller}, {Kulkarni}, {Bulla}, {Masci}, {Goldstein}, {Goobar}, {Nugent}, {Dugas}, {Blagorodnova}, {Neill}, {Rigault}, {Sollerman}, {Nordin}, {Bellm}, {Cenko}, {De}, {Dhawan}, {Feindt}, {Fremling}, {Gatkine}, {Graham}, {Graham}, {Ho}, {Hung}, {Kasliwal}, {Kupfer}, {Laher}, {Perley}, {Rusholme}, {Shupe}, {Soumagnac}, {Taggart}, {Walters}, \& {Yan}}]{yao2019}
{Yao}, Y., {Miller}, A.~A., {Kulkarni}, S.~R., {et~al.} 2019, \apj, 886, 152

\bibitem[{{Yuan} {et~al.}(2015){Yuan}, {Lidman}, {Davis}, {Childress}, {Abdalla}, {Banerji}, {Buckley-Geer}, {Carnero Rosell}, {Carollo}, {Castander}, {D'Andrea}, {Diehl}, {Cunha}, {Foley}, {Frieman}, {Glazebrook}, {Gschwend}, {Hinton}, {Jouvel}, {Kessler}, {Kim}, {King}, {Kuehn}, {Kuhlmann}, {Lewis}, {Lin}, {Martini}, {McMahon}, {Mould}, {Nichol}, {Norris}, {O'Neill}, {Ostrovski}, {Papadopoulos}, {Parkinson}, {Reed}, {Romer}, {Rooney}, {Rozo}, {Rykoff}, {Sako}, {Scalzo}, {Schmidt}, {Scolnic}, {Seymour}, {Sharp}, {Sobreira}, {Sullivan}, {Thomas}, {Tucker}, {Uddin}, {Wechsler}, {Wester}, {Wilcox}, {Zhang}, {Abbott}, {Allam}, {Bauer}, {Benoit-L{\'e}vy}, {Bertin}, {Brooks}, {Burke}, {Carrasco Kind}, {Covarrubias}, {Crocce}, {da Costa}, {DePoy}, {Desai}, {Doel}, {Eifler}, {Evrard}, {Fausti Neto}, {Flaugher}, {Fosalba}, {Gaztanaga}, {Gerdes}, {Gruen}, {Gruendl}, {Honscheid}, {James}, {Kuropatkin}, {Lahav}, {Li}, {Maia}, {Makler}, {Marshall}, {Miller}, {Miquel}, {Ogando}, {Plazas}, {Roodman}, {Sanchez}, {Scarpine},
  {Schubnell}, {Sevilla-Noarbe}, {Smith}, {Soares-Santos}, {Suchyta}, {Swanson}, {Tarle}, {Thaler}, \& {Walker}}]{yuan2015}
{Yuan}, F., {Lidman}, C., {Davis}, T.~M., {et~al.} 2015, \mnras, 452, 3047

\end{thebibliography}

\begin{appendix}
\label{app:A}





\onecolumn
\section{Approximate nearest neighbor algorithms}

Approximate nearest neighbor algorithms rapidly identify points that are \textit{close enough} to a given query, trading off a small amount of accuracy for dramatically improved search speed in high-dimensional spaces. Mathematically, we could say that a point $p_{\mathrm{ANN}}$ is an approximate nearest neighbor of the query point $q$ if we have that,
   \begin{equation}\label{eq:ann}
   d(p_{\mathrm{ANN}},q)\leq (1+\varepsilon)\,d(p_{\mathrm{NN}},q),
   \end{equation}
   where $p_{\mathrm{NN}}$ is the exact nearest neighbor and $\varepsilon$ is some small parameter that defines the extent of the approximation. This relaxation allows for a wide range of data structures and algorithms that scale much better to high-dimensional datasets, making ANN searches indispensable in modern applications such as image retrieval, recommendation systems, and, in astrophysics, matching observational data to large simulation banks.

ANN algorithms often rely on one or more of the following approaches:

\begin{itemize}
\item Random Projection: Mapping high-dimensional data to a lower-dimensional subspace using random linear transformations, while approximately preserving relative distances between data points.
\item Tree-Based Partitioning: Building multiple trees (such as randomized $k$-d trees or vantage-point trees) and searching across them to narrow down candidate points.
\item Hashing Schemes: Using locality-sensitive hashing (LSH) to group together similar data points, determined through hash collisions, into lower dimension spaces; different metrics for similarity can be specified and tailored to the application.
\item Graph-Based Structures: Creating navigable graphs of elements, where each point is connected to a subset of neighbors by directed edges, which can be iteratively explored by moving to a node in the pool that is closer to the query point until there exist no closer nodes. 
\end{itemize}

In many astrophysical contexts, dimensionality can be moderate (e.g., tens of features describing the shape, brightness, or other metrics of a light curve), but the dataset size can be extremely large—thousands or tens of thousands of simulations, each with multiple time points. An ANN approach can thus provide near-instantaneous retrieval of likely matches, even if it occasionally misses the absolute best match.

\begin{figure*}[h!]
    \centering
    {\includegraphics[width=0.9\hsize]{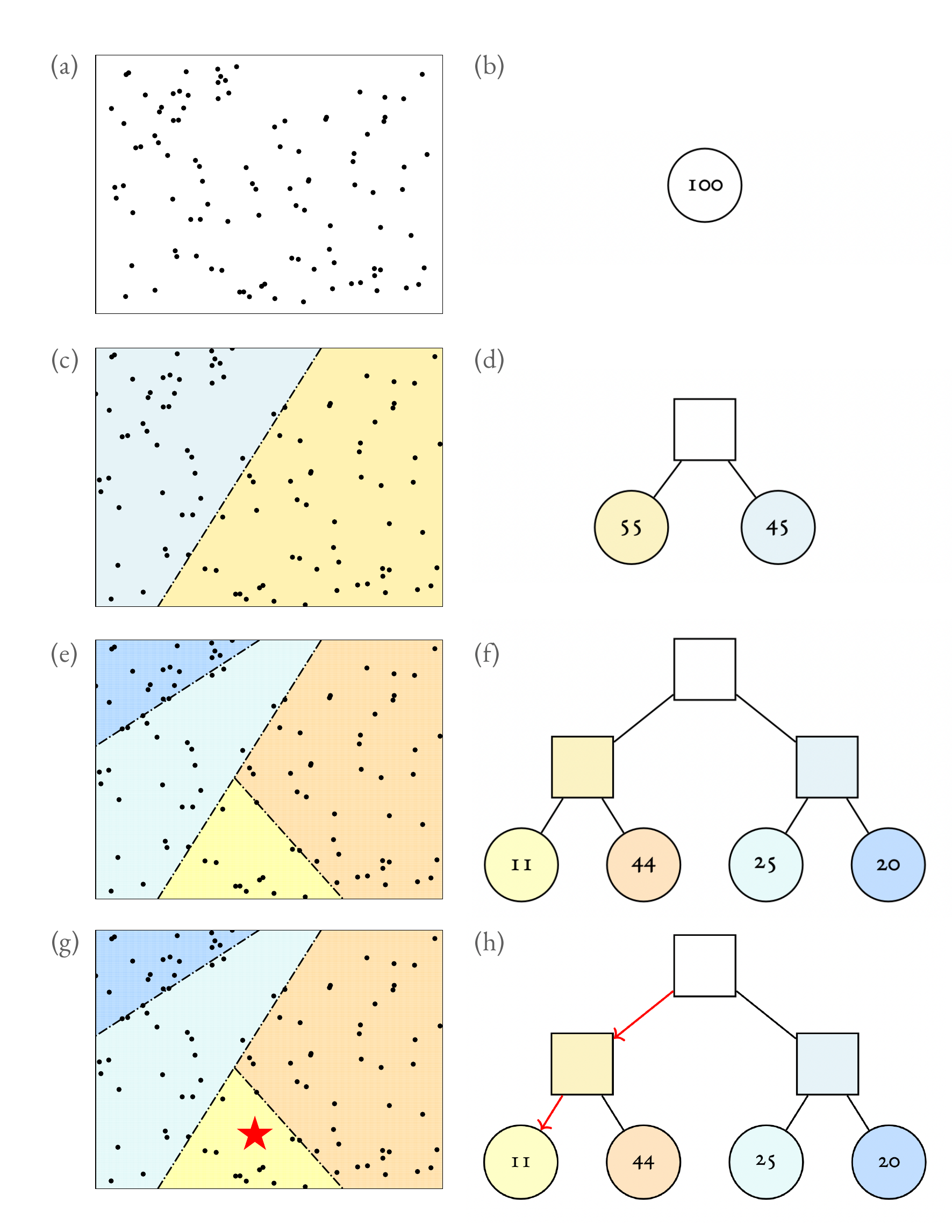}}
      \caption{Illustration of how random projections (hyperplanes) can partition data into “buckets” for approximate nearest neighbor (ANN) searches. The left column shows a two‐dimensional example in which random hyperplanes repeatedly divide the dataset, which consists of 100 points in the $x-y$ plane, into subspaces; these subspaces are then organized into a binary tree structure as shown in the right column. Note that points that lie near each other in the data space are likely to end up in the same or neighboring branches of the tree. In the final row, a query point (red star) is introduced, and the search procedure navigates the tree to find the relevant bucket of points that are most similar to the query. In practice, the dataset would typically reside in a much higher‐dimensional space, and many more random hyperplanes would be used to create finer‐grained buckets. Despite the simplified illustration here, the same principle underlies large‐scale ANN algorithms in real applications.}
      \label{fig:ch3_ANN}
\end{figure*}

Most ANN methods share a few core steps. First, a preprocessing phase builds an index structure from the dataset of interest. The point of the index data structure is is to divide the into more manageable chunks, allowing faster traversal during the search phase. This preprocessing phase can be time-consuming but is done once. Then, during query time, the algorithm uses the index structure to prune the search space, avoiding a full linear scan of all points.  For example, in a random projection-based algorithm, previously described above, the dataset is repeatedly projected onto random lower-dimensional hyperplanes. Each projection divides the data points into buckets or leaves in a tree-like structure. In the querying phase, the same projections are applied to the query point and to identify the relevant buckets or leaves. Although these methods do not guarantee that the absolute nearest neighbor is always found, they do so with high probability, and the error margin can be tuned by adjusting the number of projections or the size of each bucket. A simplified example of how the algorithm works is shown in Fig. \ref{fig:ch3_ANN}.

An important consideration is the trade-off between accuracy and speed. By allowing a small approximation error when determining data partitions such as projections or buckets, users can drastically reduce query times. Additionally, advanced data structures can be combined or stacked—multiple random trees, for instance—to refine the approximation. The user can control parameters like the number of trees, the search depth, or the allowed number of distance computations to tune the balance between query speed and result quality.

\texttt{annoy} (Approximate Nearest Neighbors Oh Yeah) is a popular C++ library with Python bindings, originally developed at Spotify for large-scale music recommendation tasks~\cite{annoy}. It is designed to handle high-dimensional data and quickly retrieve neighbors that are “close enough” to the query. It is designed in particular to use static files as indexes, so that indexes can be shared across processes, and to minimize memory footprint. Internally, \texttt{annoy} builds multiple random projection trees. At every intermediate node in the tree, a hyperplane is chosen based on two randomly selected points from the subset, which then divides the space into two subspaces. This process of partitioning the data is done recursively using random hyperplanes to reach a forest of trees (the number of times this is done represents the tradeoff between accuracy and speed, and is determined by a parameter specified by the user). In the query phase, the algorithm then traverses these trees, exploring only relevant branches to find approximate nearest neighbors. Usage of this library has both pros and cons, which we briefly discuss below. 

\texttt{annoy} offers several advantages, relevant for our purposes. Its simplicity is one of its key strengths, as users simply insert their data points, build the index, and then query for neighbors, all using the Python interface layer written over the C++ functions, classes, and objects. The library is highly memory efficient because it relies on memory mapping, which allows large datasets to be stored on disk and accessed without loading everything into RAM at once—a critical feature when dealing with massive simulation banks. The package also delivers fast build and query times; once the index is constructed, queries are executed rapidly, benefiting from the performance gains of its underlying C++ implementation. Finally, \texttt{annoy} provides a high degree of flexibility through parameter tuning by allowing users to specify the number of trees (\texttt{n\_trees}) to build when indexing the dataset and the number of nodes to explore at query time (\texttt{search\_k}), enabling users to strike a tunable balance between accuracy and speed. Increasing \texttt{n\_trees} allows for more comprehensive coverage of the data space (more random projections are considered), but also increases both memory usage and query time. Likewise, increasing \texttt{search\_k} should increase the accuracy of the algorithm by considering a larger set of potential nearest neighbors, while sacrificing query speed.

Necessarily, \texttt{annoy} also has some drawbacks. Like all ANN methods, it returns an approximate nearest neighbor rather than the exact match, and while the resulting error is typically small, it can be a concern for highly sensitive tasks. In our application, however, the relatively small number of observed light curves of interest allows for manual verification of the predicted nearest neighbor as well as the subsequent candidates, thereby ensuring that the algorithm does not yield an obviously incorrect match. Second, annoy’s index is static—once built, it cannot be updated incrementally, so any addition or removal of points requires rebuilding the index. For our purposes, this limitation is not critical because our simulation bank does not change after being generated once. Finally, annoy may be less effective for extremely high-dimensional data, as the performance of random projection techniques can degrade when the number of dimensions is very large. However, since our feature vectors for light curves are of moderate dimension, this issue is not a significant concern for our purposes.

\section{Calculation of supermassive black hole mass}
\label{app:M_SMBH}

We start with each host galaxy's redshift $z$, $g$-band magnitude, and $r$-band magnitude. The luminosity distance $D_L$ is calculated from $z$ using the \texttt{cosmology} subpackage within \texttt{Astropy}~\cite{astropy2022}, using cosmological constants from Planck 2018. The r-band luminosity $L_r$ is calculated as,
\begin{equation}
	L_r=L_{r,\odot}\times10^{-0.4\left(M_r-M_{r,\odot}\right)},
\end{equation}
where in solar units we have that $L_{r,\odot}=1$ and $M_{r,\odot}=4.67$. Over the $g-r$ range of interest $0.3\lesssim g-r\lesssim 1$, the stellar mass to luminosity ratio is given by \citet{bell2003},
\begin{equation}
	\log_{10}\left(\frac{M_*}{L_r}\right)=-0.306+1.097(g-r),
\end{equation}
allowing us to calculate the stellar mass of the host. Finally, we use the follow relation provided by \citet{reines2015} to relate the stellar mass $M_*$ to the SMBH mass $M_{\bullet}$,
\begin{equation}
	\log_{10}\left(\frac{M_{\bullet}}{M_{\odot}}\right)=\alpha+\beta\log_{10}\left(\frac{M_{*}}{10^{11}\,M_{\odot}}\right),
\end{equation}
with $\alpha=7.45\pm0.08$ and $\beta=1.05\pm0.11$ for AGN host galaxies and $\alpha=8.95\pm0.09$ and $\beta=1.40\pm0.21$ for ellipticals and classical bulges, for largely quiescent galaxies. Given the discrepancy between the two mass relations, it might seem natural to only include the latter for the purpose of our study, since we consider quiescent hosts. However, \citet{reines2015} note that the difference between the mass relations is not necessarily due to nuclear activity (or the lack thereof), as the SMBH masses themselves are derived using different methods that are known to have their own selection biases. 

\FloatBarrier 
\clearpage

\end{appendix}
\end{document}